\documentclass[aps,prb,twocolumn,floatfix,superscriptaddress]{revtex4-2}
\usepackage{psfrag,epsfig,amsfonts,amssymb,amsmath,wasysym,bm}
\usepackage{bbold}
\usepackage{xcolor}

\usepackage[inline]{enumitem}
\usepackage[hidelinks]{hyperref}

\newcommand{\<}{\langle}
\renewcommand{\>}{\rangle}
\newcommand{\ket}[1]{\lvert #1 \rangle}
\newcommand{\bra}[1]{\langle #1 \rvert}

\newcommand{\I}{\mathrm{i}}

\newcommand{\id}{\mathbb 1}

\newcommand{\CC}{{\mathbb C}}
\newcommand{\NN}{{\mathbb N}}

\newcommand{\tr}{\operatorname{tr}}
\newcommand{\hr}{{\cal H}}

\newcommand{\mc}{\mathrm{th}}
\newcommand{\eq}{\mathrm{eq}}

\newcommand{\lmat}{\left( \begin{matrix}}	
\newcommand{\rmat}{\end{matrix} \right)}

\providecommand{\opnorm}[1]{\|#1\|_{\!\!\; {\rm op}}}

\definecolor{purple}{RGB}{128, 0, 128}

\definecolor{darkgreen}{RGB}{0, 128, 0}

\begin{document}

\title{Basis dependence of eigenstate thermalization}

\author{Lennart Dabelow}
\affiliation{School of Mathematical Sciences, Queen Mary University of London, London E1 4NS, UK}

\author{Christian Eidecker-Dunkel}
\affiliation{Faculty of Physics, 
	Bielefeld University, 
	33615 Bielefeld, Germany}

\author{Peter Reimann}
\affiliation{Faculty of Physics, 
	Bielefeld University, 
	33615 Bielefeld, Germany}

\date{\today}

\begin{abstract}
Eigenstate thermalization refers to the property that an
energy eigenstate of a many-body system is indistinguishable
from a thermal equilibrium ensemble at the same energy
as far as expectation values of local observables are concerned.
In systems with degeneracies,
the choice of an energy eigenbasis is not unique
and the fraction of basis states exhibiting eigenstate thermalization can vary.
We present a simple example where this fraction vanishes in the thermodynamic limit for one basis choice,
but remains nonzero for another choice.
In other words,
the weak eigenstate thermalization hypothesis is satisfied in the first, but violated in the second basis.
We furthermore prove that
degeneracies must abound
whenever a system 
is simultaneously
symmetric under spatial translations and reflection.
Finally,
we derive general bounds on how
strongly eigenstate thermalization may depend on the
choice of the basis, and we
reveal
some
interesting implications regarding the temporal 
relaxation properties of such systems.
\end{abstract}

\maketitle

\section{Introduction}
\label{s1}

The eigenstate thermalization hypothesis (ETH) \cite{deu91, sre94, rig08}
postulates that the expectation values of 
a 
local observable $A$
in the eigenstates of a generic 
many-body Hamiltonian $H$
approach the thermal expectation values
at the corresponding energy in the thermodynamic limit.
Put differently,
an eigenstate $\ket{n}$ of $H$ with energy $E_n$ exhibits eigenstate thermalization if
$A_{nn} := \bra{n} A \ket{n} \to A_{\mc}$
as the degrees of freedom $L \to \infty$,
and the ETH conjectures that this holds for all eigenstates $\ket{n}$
(not too close to the edges of the spectrum).
Here $A_{\mc}$ is the thermal expectation value at energy $E = E_n$,
e.g.,
\begin{equation}
	A_{\mc} := \frac{1}{\lvert S \rvert} \sum_{m \in S} A_{mm}
\label{1}	
\end{equation}
in the microcanonical ensemble,
where $S$ collects all indices $m$ whose energies lie 
in a macroscopically small energy window around $E$,
and $|S|$ indicates the number of elements in $S$.

This postulate has been verified in numerous numerical examples 
(see, for instance,
Refs.~\cite{rig08, gen12, sor14, kho15, mon16})
and is generally believed to explain why generic many-body quantum systems thermalize,
meaning that the expectation values of local observables
become indistinguishable from the thermal equilibrium values at late times.
However, it still amounts to an unproven hypothesis and there are several examples of 
mechanisms that may
lead to violations of this (strong) ETH, such as integrability \cite{ess16, vid16}, many-body 
localization \cite{nan15, sie25}, quantum many-body scars 
\cite{ser21, mou22, cha23}, and Hilbert space fragmentation \cite{mou22, cha23}.

A weaker formulation of the ETH
only asserts 
that 
the variance
\begin{equation}
\label{2}
	\Delta^2_A
		:= \frac{1}{\lvert S \rvert} \sum_{n \in S} \left( A_{nn} - A_{\mc} \right)^2 
\end{equation}
of the diagonal matrix elements of 
a 
local observable $A$
vanishes in the thermodynamic limit $L \to \infty$.
This 
so-called weak ETH
is provably true for a large class of systems exhibiting translational invariance 
\cite{bir10, mor16, iyo17, mor18, kuw20, kuw20a}.
Moreover, even though the weak ETH is insufficient for thermalization in general,
it can be used to demonstrate return to equilibrium after local quenches under 
certain additional assumptions \cite{far17, dab22}.

The weak ETH still leaves room
for the existence of non-thermal eigenstates.
However, 
their 
relative number
is bound to vanish in the thermodynamic limit,
i.e., the weak ETH enforces eigenstate thermalization at least for {\em most} 
eigenstates
\cite{bir10, mor16, iyo17, mor18, kuw20, kuw20a}.
In turn, a violation of the weak ETH means
\cite{foot1}
that a non-vanishing fraction of
eigenstates must be non-thermal
(at least under the usual tacit assumption that local 
observables are bounded operators).

We 
also
remark that the above-mentioned proofs of the weak 
ETH \cite{bir10, mor16, iyo17, mor18, kuw20, kuw20a}
are valid independently of whether the considered system is 
integrable or not (even translationally 
invariant systems consisting of non-interacting subsystems are admitted).
Furthermore, and in contrast to the conventional (strong) ETH, 
neither 
the edges of the spectrum
nor 
additional conserved quantities 
play any special role.

More precisely speaking,
the proofs of the weak ETH demonstrate that,
for any
translationally invariant 
Hamiltonian $H$
with local interactions,
\emph{there exists} an energy eigenbasis $\{ \ket{n} \}$
such that $\Delta^2_A \to 0$ as $L \to \infty$
\cite{bir10, mor16, iyo17, mor18, kuw20, kuw20a}.
This eigenbasis is unique (up to trivial phase factors)
if the spectrum of $H$ is nondegenerate.
If there are degeneracies, however,
the variance $\Delta^2_A$ in~\eqref{2} 
generally becomes
basis dependent.

In this work,
we address this basis dependence of the weak ETH.
In particular,
(i) 
we demonstrate that the 
spectrum of a translationally invariant Hamiltonian 
must exhibit a large number of degeneracies in the most
commonly studied cases obeying a spatial reflection symmetry;
(ii) we identify
the
specific energy eigenbases in which $\Delta^2_A$ (and thus any potential 
ETH violation) are minimal and maximal, respectively;
(iii) we provide an example of a system that satisfies the weak ETH in 
one basis, but violates it in another one;
(iv) we discuss various interesting implications of these findings 
for equilibration, thermalization, and dynamical auto-correlation functions.

Since the property of whether or not a system thermalizes
cannot depend on the energy eigenbasis used to evaluate the ETH,
our findings pinpoint a subtle, but important problem with the interpretation 
of eigenstate thermalization as the mechanism behind thermalization in 
systems with degeneracies.
Fundamentally speaking, this may be considered as the main 
message of our present paper.
From a practical point of view,
our findings 
imply
that numerical 
studies
of the ETH which exploit symmetries in the presence of degeneracies 
may lead to physically wrong conclusions.

\section{Degeneracies in translationally invariant systems}
\label{s2}

For the sake of simplicity,
we focus in this section on one-dimensional lattice systems
(generalizations are straightforward),
where
every lattice site is associated with a copy of the local, $D$-dimensional 
Hilbert space $\hr_{\mathrm{loc}}$ with a local  orthonormal 
basis  (ONB) $\{ \ket{\sigma} \}_{\sigma=0}^{D-1}$.
The total Hilbert space is $\hr = \hr_{\mathrm{loc}}^{\otimes L}$,
and we denote the $D^L$ states of the associated 
product ONB by 
\begin{eqnarray}
\ket{\bm\sigma} \equiv \ket{ \sigma_1 \cdots \sigma_L } := 
\bigotimes_{l=1}^L\,  \ket{\sigma_l}
\ .
\label{3}
\end{eqnarray}
The translation operator $T$ is defined as usual via its action on the product states,
\begin{eqnarray}
T \ket{ \sigma_1 \sigma_2 \cdots \sigma_{L-1} \sigma_L } 
:= 
\ket{ \sigma_L \sigma_1 \sigma_2 \cdots \sigma_{L-1} }
\ .
\label{4}
\end{eqnarray}
One readily infers that $T$ is a unitary operator and that $T^L=\id$ (identity operator).

Let us now consider a Hamiltonian $H$ that is translationally invariant,
i.e., it commutes with $T$.
Hence, there exists a common eigenbasis $\{ \ket{n} \}$ of $H$ and $T$,
meaning that $H \ket{n} = E_n \ket{n}$ and $T \ket{n} = \lambda_n \ket{n}$ for all $n$.
Furthermore, defining
\begin{eqnarray}
\gamma_\nu := e^{2\pi i \nu/L}
\label{5}
\end{eqnarray}
for $\nu\in\{0,...,L-1\}$,
and recalling that $T^L = \id$,
it follows that
$\lambda_n\in\{\gamma_0,...,\gamma_{L-1}\}$ for all $n$
and thus $T^{L-1} \ket{n} = \lambda_n^* \ket{n}$.

The salient point is that nearly all commonly considered Hamiltonians $H$
which are invariant under $T$
are also invariant under the action of the parity (spatial reflection) operator $P$
with $P \ket{\sigma_1 \cdots \sigma_L} := \ket{ \sigma_L \cdots \sigma_1}$.
Common examples are Heisenberg- or Ising-type models.

Assuming this additional symmetry $[H, P] = 0$,
we define $\ket{n}' := P \ket{n}$ and observe that $H \ket{n}' = E_n \ket{n}'$.
One readily verifies $T P \ket{\bm\sigma} = P T^{L-1} \ket{\bm\sigma}$
for all $ \ket{\bm\sigma}$, which implies $TP=PT^{L-1}$ and thus
$T \ket{n}' = T P \ket{n} = P T^{L-1} \ket{n} = \lambda_n^* \ket{n}'$.
Therefore, $\ket{n}'$ must be linearly independent of $\ket{n}$ whenever $\lambda_n^* \neq \lambda_n$,
resulting in a degeneracy of the energy eigenvalue $E_n$.
Furthermore,
assuming $\lambda_n^* = \lambda_n$ for all $n$ would imply $\lambda_n^2 = 1$ and thus $T^2 = \id$,
which is obviously violated  if $L > 2$.
Consequently, many-body Hamiltonians that simultaneously commute 
with $T$ and $P$ must exhibit at least one degeneracy.

The natural next step is to show that there exist more 
than one degeneracy.
To this end, we recall that 
all eigenvalues of $T$ are of the form
(\ref{5}) for some
$\nu\in\{0,...,L-1\}$.
Furthermore, for an arbitrary but fixed $\nu\in\{0,...,L-1\}$,
we denote by $\hr_\nu$ the eigenspace of $T$
with eigenvalue $\gamma_\nu$.
Finally, we show in Appendix~\ref{app1} that
the dimension of each $\hr_\nu$ can 
be lower bounded as 
\begin{eqnarray}
\mbox{Dim}(\hr_\nu)\geq
\frac{D^L}{L}  
\left( 1 - \frac{\ln L}{ D^{L/2} \ln 2} \right) 
\label{6}	
\end{eqnarray}
for all $\nu\in\{0,...,L-1\}$.
Observing that $\mbox{Dim}(\hr_\nu)$
equals $D^L-\sum_{\mu\not=\nu}\mbox{Dim}(\hr_\mu)$,
and exploiting (\ref{6}) for each summand $\mbox{Dim}(\hr_\mu)$
also yields an immediate upper bound for 
$\mbox{Dim}(\hr_\nu)$.
For large $L$, those lower and upper bounds imply that
the $D^L$ simultaneous eigenvectors 
$|n\rangle$ of  $H$ and $T$ are 
nearly equally distributed 
among the $L$ different eigenspaces $\hr_\nu$ of $T$
(as one possibly might have intuitively expected).
We also note that the derivation of this result in Appendix~\ref{app1}
is valid for any translationally invariant Hamiltonian $H$ 
(independently of whether it commutes with $P$ or not).

Eq.~(\ref{5}) implies that $\gamma_\nu\not=\gamma_\nu^\ast$ unless $\nu\in\{0,L/2\}$
(the second option, $\nu=L/2$,  is only possible for even $L$).
It follows that there are at least $(L-2)$ eigenspaces $\hr_\nu$
with the property $\gamma_\nu\not=\gamma_\nu^\ast$.
All eigenvectors $|n\rangle$ contained in those eigenspaces
thus fulfill the above mentioned degeneracy
criterion $\lambda_n^\ast\not=\lambda_n$.
Exploiting Eq.~(\ref{6}),
the total number $N_{\rm deg}$ of degenerate 
energy eigenvalues can thus be lower bounded as
\begin{equation}
N_{\rm deg} 
\geq
D^L \left( 1 - \frac{2}{L} \right) \left( 1 - \frac{\ln L}{ D^{L/2} \ln 2} \right) .
\label{7}	
\end{equation}
In other words, among the $D^L$ eigenvalues of 
a translational and reflection symmetric Hamiltonian,
the fraction $N_{\rm deg}/D^L$ of degenerate eigenvalues approaches 
unity in the thermodynamic limit $L\to\infty$, while the fraction of 
\emph{non}degenerate eigenvalues approaches zero.
Note that all those degeneracies are predicted to be at least two-fold,
while no statement about possibly existing higher degeneracies 
is made.

This is the first main result of our paper: 
Commonly studied translationally invariant systems
are usually at the same time also invariant under spatial 
reflection. If so, the vast majority of the energy eigenvalues must 
exhibit degeneracies.

\section{Minimally and maximally ETH-violating bases}
\label{s3}

Let us now return to the general setup
and consider a possibly degenerate Hamiltonian $H$.
The nature of the Hamiltonian and the source of degeneracies 
(for instance symmetries)
are irrelevant for the discussion in this section.

\subsection{Main formal result}
\label{s31}

To begin with, we note that the ETH quantifier in Eq.~(\ref{2}) can be rewritten
by means of Eq.~(\ref{1}) as
\begin{equation}
	\Delta^2_A
         = \left(\frac{1}{\lvert S \rvert} \sum_{n \in S} A^2_{nn}\right) - A^2_{\mc}
         = \frac{1}{\lvert S \rvert} \sum_{n \in S} (A^2_{nn} - A^2_{\mc})
         \, .
\label{8}
\end{equation}
Next, we denote by 
$W$ the set of all
distinct energy eigenvalues in the microcanonical window $S$.
Furthermore, we define $g_E$ as the multiplicity of $E \in W$,
i.e., the dimension of the eigenspace of energy $E$.
Let $\{ \ket{E, \nu} \}_{\nu=1}^{g_E}$ be an 
orthonormal basis
 of that eigenspace with energy $E$.
For this basis,
the ETH quantifier~\eqref{8} can be decomposed
as
\begin{eqnarray}
	\Delta_A^2 
	&=& 
	\frac{1}{\lvert S \rvert} \sum_{E \in W} g_E \, \Delta_{A, E}^2 \,, 
\label{9}
\\
\Delta_{A,E}^2 
&:=&
\frac{1}{g_E} \sum_{\nu = 1}^{g_E} \left(A^2_{E,\nu\nu} - A^2_{\mc}\right)
\nonumber
\\	
& = &
\frac{1}{g_E} \left(\sum_{\nu = 1}^{g_E} A^2_{E,\nu\nu}\right) - A_{\mc}^2
\label{10}
\end{eqnarray}
with $A_{E,\mu\nu} := \bra{E, \mu} A \ket{E, \nu}$.
To find the minimal (maximal) possible value of $\Delta_A^2$,
it suffices to minimize (maximize) $\Delta_{A,E}^2$ for every $E$ individually.

We first inspect the maximum.
Denoting the projector onto the energy eigenspace of $E$ by $P_E$, we 
observe that
\begin{eqnarray}
	\tr[(P_E A)^2]
		&= & \sum_{\mu, \nu = 1}^{g_E} A_{E,\mu\nu} A_{E,\nu\mu} 
		\nonumber
		\\
		&= & g_E (\Delta_{A,E}^2 + A_{\mc}^2) + \sum_{\mu \neq \nu} \lvert A_{E,\mu\nu} \rvert^2 
		\, ,
		\label{11}
\end{eqnarray}
where the last sum runs over all $\mu,\nu\in\{1,...,g_E\}$ with $\mu\not=\nu$.
Next, we focus on a fixed subspace with energy $E \in W$ and
 $g_E > 1$, and
 introduce a second orthonormal basis $\{ \ket{E, \nu}' \}_{\nu=1}^{g_E}$ 
with matrix elements $A_{E,\mu\nu}'$ of $A$ and associated ETH quantifier $\Delta_{A,E}^{\prime 2}$.
Since the trace on the left-hand side of  Eq.~(\ref{11})
is basis independent, the same
expression on the right-hand side holds analogously for the 
second basis $\{ \ket{E, \nu}' \}$, implying that
\begin{equation}
	\Delta_{A,E}^2 + \frac{1}{g_E} \sum_{\mu\neq\nu} \lvert A_{E,\mu\nu} \rvert^2
	= \Delta_{A,E}^{\prime 2} + \frac{1}{g_E} \sum_{\mu\neq\nu} \lvert A_{E,\mu\nu}' \rvert^2 \,.
\label{12}	
\end{equation}
If we now choose one of the bases,
say $\{ \ket{E, \nu}' \}$,
such that $A_{E,\mu\nu}' = 0$ whenever $\mu \neq \nu$
(i.e., we diagonalize $A$ within the eigenspace of energy $E$),
then
\begin{equation}
	\Delta_{A,E}^2 + \frac{1}{g_E} \sum_{\mu\neq\nu} \lvert A_{E,\mu\nu} \rvert^2
	= \Delta_{A,E}^{\prime 2}
\label{13}	
\end{equation}
and consequently
\begin{equation}
	\Delta_{A,E}^2 \leq \Delta_{A,E}^{\prime 2}
		= \frac{ \tr[ (P_E A)^2 ]}{g_E} - A_{\mc}^2 \,.
\label{14}	
\end{equation}

We thus arrive at the second main result of our paper:
The energy eigenbasis where $A$ is diagonal in every degenerate 
energy subspace maximizes the ETH quantifier
(\ref{10})
in every subspace
and thus, via~\eqref{9}, also over the entire microcanonical energy window.
In other words,
this is the most ``dangerous'' basis with regard to violations of the weak ETH.

Aiming also for the opposite extreme of the basis that minimizes the ETH violation,
we return to Eq.~\eqref{10} and consider
\begin{equation}
\label{eq:ETHVio:LowerBound}
\begin{aligned}
	\sum_{\nu=1}^{g_E} A_{E,\nu\nu}^2
		&= \sum_{\nu=1}^{g_E} \left( A_{E,\nu\nu} - \frac{ \tr(P_E A) }{ g_E } \right)^2 + \frac{ [ \tr(P_E A) ]^2 }{ g_E } \\
		&\geq \frac{ [ \tr(P_E A) ]^2 }{ g_E } \,.
\end{aligned}
\end{equation}
Since the last term is basis independent,
the relation~\eqref{eq:ETHVio:LowerBound} gives a lower bound for the sum of squared diagonal matrix elements in any basis.
Substituted into Eq.~\eqref{10},
it yields
\begin{equation}
	\Delta_{A,E}^2 \geq \frac{ [ \tr(P_E A) ]^2 }{ g_E } - A_{\mc}^2 \,.
\end{equation}

Furthermore,
the bound is saturated if $A_{E,\nu\nu} = \tr(P_E A) / g_E$ for all $\nu$.
The smallest ETH violation for a given observable $A$ is thus obtained for the basis where all diagonal matrix elements of $A$ assume the same value within degenerate energy subspaces (preserving the trace).
Such a basis always exists
because, as shown in Example~2.2.3 of Ref.~\cite{hor13},
every complex matrix is unitarily similar to a matrix whose diagonal entries are equal.

In summary,
the basis dependence of the ETH quantifier $\Delta_A^2$ from~\eqref{2}
varies within the range
\begin{equation}
	\frac{1}{\lvert S \rvert} \sum_{E \in W} [\tr(P_E A)]^2 
	- A_{\mc}^2 \leq \Delta_A^2 \leq \frac{1}{\lvert S \rvert} \sum_{E \in W} \!\! \tr[(P_E A)^2] - A_{\mc}^2 \,.
\label{17a}
\end{equation}
The lower bound is assumed by choosing a basis with constant diagonal 
matrix elements for $A$ in every degenerate energy subspace.
The upper bound is assumed by choosing a basis such that $A$ 
is diagonal in every such subspace.
Note that generally those extremizing bases will 
depend on the observable $A$.

Since physical system properties cannot depend on the basis choice,
conclusions about a system's behavior can only be drawn from the appropriate bound.
In particular, a basis-independent weak ETH should instead be formulated as a vanishing upper bound in~\eqref{17a}, i.e.,
\begin{equation}
\label{17b}
	\hat{\Delta}_A^2 := \frac{1}{\lvert S \rvert} \sum_{E \in W} \!\! \tr[(P_E A)^2] - A_{\mc}^2 \to 0
\end{equation}
in the thermodynamic limit.
We will come back to concrete physical implications of this in Sec.~\ref{s6}.

\subsection{Generalization}
\label{s32}

The line of reasoning from the previous section can be
readily generalized as follows:
Consider an arbitrary ``equilibrium ensemble'' of the form
\begin{eqnarray}
\rho_{\rm eq}:= f(H)
\label{15}
\end{eqnarray}
with any function $f$ such that $\rho_{\rm eq}$ is a well-defined
density operator, i.e., $\rho_{\rm eq}\geq 0$ and 
$\tr\{\rho_{\rm eq}\}=1$.
By definition, for any given orthonormal eigenbasis $|n\rangle$ of $H$ with corresponding
eigenvalues $E_n$, Eq.~(\ref{15}) thus assumes the form
\begin{eqnarray}
\rho_{\rm eq} & = & \sum_n p_n | n\rangle \langle n|
\label{16}
\end{eqnarray}
with the ``level populations'' $p_n :=  f(E_n)$. 
It follows that
$\rho_{\rm eq}$ commutes with $H$
and therefore is a steady state 
(hence the name {\em equilibrium ensemble}).
Moreover, it may be viewed as a so-called {\em diagonal ensemble}.
Particularly important examples are the
microcanonical and the canonical ensembles.

\begin{figure*}
	\includegraphics[scale=1]{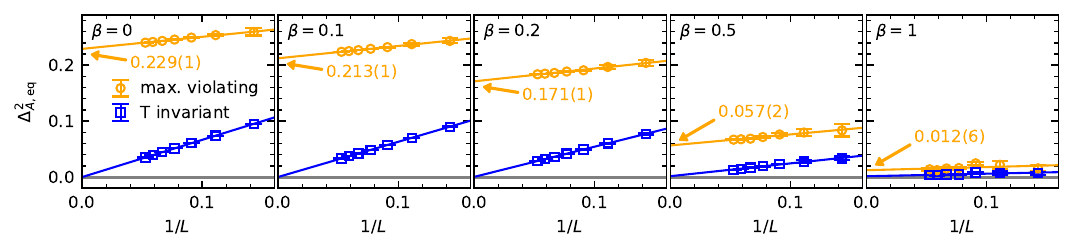}
	\caption{ETH quantifier $\Delta_{A,\eq}^2$ from~\eqref{18} vs.\ $L^{-1}$ for the canonical ensemble~\eqref{eq:rhoCan}, Hamiltonian $H$ from~\eqref{55}, and observable $A = s_1^z$, for two basis choices:
		one where $A$ is diagonal in all energy eigenspaces (maximal ETH violation, orange)
		and a translationally invariant one (provably satisfying weak ETH, blue).
		Panels correspond to different values of the inverse temperature $\beta$.
		Symbols are numerical results obtained via dynamical typicality with error bars indicating the 
		corresponding standard error (cf.\ Appendix~\ref{app:DynTyp}; most error bars are actually 
		not discernible on the scale of these plots); lines are linear fits and orange numbers the 
		resulting extrapolation of $\Delta_{A,\eq}^2$ to $L \to \infty$ ($y$-axis intercept, including, 
		in parentheses, the uncertainty in the last digit).}
	\label{fig:ETHVio}
\end{figure*}

Similarly as in Eq.~(\ref{1}), the corresponding equilibrium value
of an observable $A$ is defined as
\begin{eqnarray}
A_{\rm eq}:= \tr\{\rho_{\rm eq} A\}=\sum_n p_n A_{nn}
\ ,
\label{17}
\end{eqnarray}
while the ETH quantifier (\ref{2}) now assumes the
form
\begin{equation}
\Delta^2_{A,{\rm eq}}:= \sum_{n} p_n (A_{nn}-A_{\rm eq})^2 
\ .
\label{18}
\end{equation}
By means of a straightforward modification of the arguments
in the previous section,
this quantity can again be shown
to assume its maximal (minimal)
value when choosing the basis $|n\rangle$ 
such that $A$ is diagonal (has constant diagonal matrix elements) in every eigenspace of $H$.
The basis-independent upper bound analogous to~\eqref{17b} is then given by
\begin{equation}
\label{18b}
	\hat{\Delta}_{A, {\rm eq}}^2 := \sum_{E} f(E) \tr[(P_E A)^2] - A_{\rm eq}^2 \,,
\end{equation}
where the sum runs over all
pairwise distinct eigenvalues 
of $H$.

\section{Example for basis dependence of the weak ETH}
\label{s5}

To illustrate 
the 
basis dependence of weak eigenstate thermalization,
we consider 
the
spin-1 Hamiltonian with periodic boundary conditions
\begin{equation}
	\label{55}
	H = - \sum_{l=1}^L \left[ s_l^+ (s_{l+1}^-)^2 s_{l+2}^+ + s_l^- (s_{l+1}^+)^2 s_{l+2}^- \right] ,
\end{equation}
where $s_l^\pm := s_l^x \pm \I s_l^y$,
$s_l^a$ ($a = x, y, z$) are the standard spin-1 operators acting 
on site $l$, and $s_{l+L}^a:=s_{l}^a$.

This model plays a prominent role in Ref.~\cite{sal20}, where it
has been shown to be magnetization- and dipole-moment-conserving
(see also Appendix~\ref{app:DynTyp}),
and -- most importantly -- to exhibit Hilbert space fragmentation.
More precisely speaking, the main focus in Ref.~\cite{sal20} was on
the correspondingly modified Hamiltonian~\eqref{55}  with open boundary 
conditions, while we will choose periodic boundary conditions in the
following.
This modification is
essential
since it
ensures that the results from Refs.~\cite{bir10, mor16, iyo17, mor18, kuw20, kuw20a} 
are applicable, guaranteeing 
the existence of 
at least one energy eigenbasis in which the weak ETH holds,
namely 
any 
simultaneous eigenbasis of $H$ and $T$.

We focus on the single-site magnetization $A = s_1^z$ and the ETH quantifier 
$\Delta_{A,\eq}^2$ from Eq.~\eqref{18}
for the canonical ensemble (Gibbs state),
\begin{equation}
	\label{eq:rhoCan}
	p_n = Z^{-1} e^{-\beta E_n}
\end{equation}
with $Z := \sum_n e^{-\beta E_n}$.
Bearing in mind Sec.~\ref{s32} and the very generally 
established
equivalence of the canonical and microcanonical ensembles in sufficiently large 
systems \cite{tou15, bra15, tas18, kuw20, kuw20a},
it is expected that the canonical $\Delta_{A,\eq}^2$ from Eq.~(\ref{18}) and its microcanonical 
counterpart $\Delta_{A}^2$ from~\eqref{2}
show the same scaling behavior with the degrees of freedom $L$ in a fixed basis.
A more detailed justification will be provided in Sec.~\ref{s42}.

In Fig.~\ref{fig:ETHVio},
we show $\Delta_{A,\eq}^2$ as a function of $1/L$
for two different choices of the energy eigenbasis
and system sizes $L = 7, 9, \ldots, 19$.
Details on how $\Delta_{A,\eq}^2$ was evaluated numerically
using 
dynamical typicality
methods \cite{bar09, rei20} are provided in Appendix~\ref{app:DynTyp}.

The first basis choice (blue squares in Fig.~\ref{fig:ETHVio}) consists of 
translationally invariant eigenstates $\ket{n}$.
In this case,
the results from Refs.~\cite{bir10, mor16, iyo17, mor18, kuw20, kuw20a} rigorously 
demonstrate the validity of the weak ETH.
In combination with the above mentioned equivalence of the ensembles,
$\Delta_{A,\eq}^2$ is thus
guaranteed to approach zero as $L \to \infty$
for any choice of the inverse temperature $\beta$.
This is confirmed by our numerics upon extrapolating the finite-size results (blue lines):
Fitting the numerical data to a linear
function
$\Delta_{A,\eq}^2 = a + b/L$
yields values $\lvert a \rvert \leq 0.002$ (with uncertainties less than $0.002$) for 
all $\beta$ values shown in the figure.

The second basis choice (orange circles in Fig.~\ref{fig:ETHVio}) 
is the maximally ETH-violating basis identified in Sec.~\ref{s3},
meaning that the observable $A = s_1^z$ is diagonal in every degenerate energy subspace.
In this basis, we observe that the ETH quantifier $\Delta_{A,\eq}^2$ does \emph{not} generally decay to zero
as the system size is increased.
Extrapolation to $L \to \infty$ (orange lines) instead yields a positive, $\beta$-dependent value
as quoted in each of the figure panels.
Only for 
asymptotically low temperatures (large $\beta$)
$\Delta_{A,\eq}^2$ vanishes (and almost uniformly so in the system size $L$).

From a practical point of view, this 
amounts to the most important result of our present paper:
an explicit example where the weak ETH is violated in one basis,
but satisfied in another one.

\section{Analytical approach}
\label{s4}

In the example from the previous section,
we inspected the ETH quantifier $\Delta_{A,\eq}^2$ for the canonical 
ensemble~\eqref{eq:rhoCan}.
By tuning the inverse temperature $\beta$,
we thus can effectively select an energy window around $E = \tr\{\rho_{\eq} H\}$,
and 
by varying $\beta$ we can
scan the entire spectrum of the Hamiltonian
for ETH violations.
In this section,
we provide a more rigorous treatment of this connection between energy windows
in the canonical and microcanonical formalisms.

\vspace*{0.9cm}

\subsection{A simple special case}
\label{s41}

In this subsection, we restrict ourselves to Hamiltonians $H$ and 
local observables $A$ which satisfy several 
simplifying assumptions,
and we show that these assumptions are sufficient
to imply a violation of the weak ETH.
An important example
of this kind 
is the setup from Sec.~\ref{s5} above.

We consider a system 
of large but finite size 
$L$ and our first assumption is that it exhibits 
a finite number $N$ of energies $\{E_n\}_{n=1}^N$.
Typical examples are spin-lattice models
such as in Eq.~(\ref{55}),
giving rise to an exponential growth of $N$ with $L$.
Second, we focus on cases where $A_{\rm th}=0$
for any microcanonical energy window with
concomitant subset $S\subset\{1,...,N\}$
in Eq.~(\ref{1}).
Third, we assume that the local observable $A$ 
exhibits a finite range of possible measurement 
outcomes (eigenvalues of $A$). More precisely 
speaking the operator norm $\opnorm{A}$
(largest eigenvalue in modulus of $A$) 
is assumed to remain bounded, say $\opnorm{A}<a$, 
for all $L$.
It follows that
\begin{eqnarray}
A^2_{nn}\leq a^2
\label{26}
\end{eqnarray}
independently of $n$ and $L$.
Finally, we assume that the quantity
\begin{equation}
\label{27}
	\Delta^2_{A,\rm{tot}}
		:= \frac{1}{N} \sum_{n=1}^N  A_{nn}^2 
\end{equation}
exhibits the property
\begin{eqnarray}
\Delta^2_{A,\rm{tot}} \geq b >0
\ .
\label{28}
\end{eqnarray}
for all $L$. One thus can infer that $a^2\geq b$.

We remark that while Eq.~(\ref{2}) (with $A_{\rm th}=0$)
may be viewed as an average over a relatively small 
microcanonical subset $S$,
Eq.~(\ref{27}) represents the corresponding 
average over the full set $\{1,...,N\}$.
Hence, Eq.~(\ref{28}) amounts to yet another variety of ETH violation,
slightly different from the previously discussed violation
of the strong or weak ETH. 
Incidentally, $\Delta^2_{A,\rm{tot}}$ in Eq.~(\ref{27}) also happens to 
coincide with the ETH quantifier 
$\Delta_{A,\eq}^2$ from Eq.~(\ref{18}) for the canonical 
ensemble (\ref{eq:rhoCan}) with inverse temperature $\beta = 0$.

Let $S_1, \ldots, S_M$ be a finite partition of $\{1,...,N\}$, i.e.,
$S_1 \cup \ldots \cup S_M=\{1,...,N\}$ and $S_i \cap S_j = \emptyset$ if $i \neq j$. 
Furthermore, we denote by  $N_b$ the number of all indices $n\in\{1,...,N\}$
with the property $A_{nn}^2 \geq b/2$.
Together with (\ref{26})-(\ref{28}) it follows that
\begin{eqnarray}
b\leq  \frac{1}{N} \sum_{n=1}^N  A_{nn}^2 \leq \frac{1}{N} \left(N_b a^2+ (N-N_b) \frac{b}{2}\right)
\label{29}
\end{eqnarray}
and thus
\begin{eqnarray}
\frac{N_b}{N}\geq \frac{b}{2a^2-b}\geq \frac{b}{2a^2}
\ ,
\label{30}
\end{eqnarray}
where we exploited that $a^2\geq b>0$ in the last step
(see in and below Eq.~(\ref{28})).
Next, we denote by $\nu_m$ the number of all indices $n\in S_m$ with the 
property $A_{nn}^2\geq b/2$.
It readily follows that $\nu_m/|S_m|\geq N_b/N$ for at least one $m\in\{1,...,M\}$.
Focusing on such a particular $m$ value we thus obtain
\begin{eqnarray}
\frac{1}{|S_m|}\sum_{n\in S_m}A_{nn}^2
\geq 
\frac{\nu_m}{|S_m|} \frac{b}{2}\geq \frac{N_b}{N} \frac{b}{2}\geq  
\frac{b^2}{4a^2}
\ .
\label{31}
\end{eqnarray}
Observing that the right-hand side is positive and independent of $L$,
and assuming that the partition of $\{1,...,N\}$ has been chosen such that each
$S_m$ corresponds to a microcanonical energy window as 
specified below Eq.~(\ref{1}),
it follows that the variance in Eq.~(\ref{2}) is lower bounded by $b^2/4a^2>0$
for at least one microcanonical energy window, and thus the weak ETH is
violated.

In summary, for setups satisfying the three assumptions from the beginning of this subsection,
a nonzero value of $\Delta_{A,\mathrm{tot}}^2$ (tantamount to an ETH violation over the entire spectrum)
implies a violation of the weak ETH in its standard formulation for microcanonical energy windows 
\cite{bir10, mor16, iyo17, mor18, kuw20, kuw20a}.
Notably, this applies to the example from 
the leftmost panel in Fig.~\ref{fig:ETHVio} because $\Delta_{A,\mathrm{tot}}^2$ 
is equal to $\Delta_{A,\eq}^2$ for $\beta = 0$ (see below Eq.~(\ref{28})).

\subsection{More general treatment}
\label{s42}

As seen in Sec.~\ref{s1}, the main focus in the context of the ETH
-- and of statistical physics in general -- 
is on large system sizes $L$.
It is therefore tacitly taken for granted that the 
considered Hamiltonians, states, and observables 
are such that the resulting properties of actual interest exhibit 
a well-defined thermodynamic limit $L\to\infty$.
Most importantly, instead of energies $E$, one thus works with
the corresponding energy densities 
\begin{eqnarray}
e:=(E-{\cal E})/L
\, ,
\label{32}
\end{eqnarray}
where ${\cal E}$ is some
physically reasonable (and generally $L$-dependent) reference
energy, for instance the ground state energy, or the average over all 
energy eigenvalues $\{E_n\}_{n=1}^N$ 
in cases where $N$ is finite (see above Eq.~({24})).
In particular, the 
counterparts of the energy eigenvalues $E_n$ are given by 
\begin{eqnarray}
e_n:=(E_n-{\cal E})/L
\, .
\label{33}
\end{eqnarray}

Along these lines, the definitions in and around Eq.~(\ref{1})
are modified as follows:
Given some energy density $e$, the set 
$S$
consists of all indices $m$ with the property that 
$e_m\in[e,e+\delta]$, where $\delta$
is macroscopically small
and microscopically large,
meaning that it approaches zero for $L\to\infty$
in such a way that the number $|S|$ of elements in
$S$ tends to infinity.
In other words, the corresponding quantity 
$A_{\rm th}$
in Eq.~(\ref{1})
is understood (and formally defined) analogously as 
when going over, e.g., from classical point particles to (local) 
particle densities, and it is taken for granted that such 
a ``local density'' of the $A_{mm}$'s in (\ref{1}) 
is well-defined
and largely independent of any further details of 
how $\delta$ approaches zero for large $L$.
In particular, 
$A_{\rm th}=A_{\rm th}(e)$
will be a practically
$L$- and $\delta$-independent 
function of $e$ for sufficiently 
large $L$.
Again in other words, this is the standard definition
in statistical mechanics of the microcanonical expectation 
value of $A$ for a system with energy density $e$,
and as such can and will be considered as ``well-defined''
from now on.

As a consequence, upon considering the auxiliary quantities 
$\tilde A_{nn}:=A_{nn}-A_{\rm th}(e_n)$ instead of $A_{nn}$, 
the second assumption above Eq.~(\ref{26}) takes the form 
$\tilde A_{\rm th}=0$, and is readily seen to be fulfilled
for asymptotically large $L$.
Likewise, one sees that the ETH quantifier for $A_{nn}$ 
in  Eq.~(\ref{2}) becomes asymptotically equal to 
the ETH quantifier for $\tilde A_{nn}$.
Hence, the line of reasoning in Sec.~\ref{s41} 
can be adapted such that it remains valid even 
without the second assumption above Eq.~(\ref{26}). 
The final conclusion is that 
the weak ETH must be violated whenever
\begin{eqnarray}
\frac{1}{N} \sum_{n=1}^N  [A_{nn}-A_{\rm th}(e_n)]^2 \geq b >0
\ .
\label{34}
\end{eqnarray}
for all sufficiently large $L$.

By means of analogous arguments as in our
above considerations regarding 
$A_{\rm th}=A_{\rm th}(e)$,
Eq.~(\ref{2}) 
can be
rewritten as
\begin{eqnarray}
\Delta_A^2(e) & = & \frac{1}{|S|}\sum_{n\in S} 
\left[A_{nn}-A_{\rm th}(e_n)\right]^2
\ ,
\label{35}
\end{eqnarray}
and it is again taken for granted that  
this quantity exhibits a well-defined
thermodynamic limit and thus will be a practically 
$L$- and $\delta$-independent 
function of $e$ for sufficiently 
large $L$.
Accordingly, the weak ETH (see below Eq.~({2})) now
requires that $\Delta_A^2(e)$ approaches zero for large $L$.
More precisely speaking, this must be the case for any
given local observable $A$ and any given energy density 
$e$.
In turn, this implies that in order to show a violation
of the weak ETH (and thus also of the strong ETH)
it is sufficient that  $\Delta_A^2(e)$ approaches
a non-vanishing large-$L$ limit for at least one 
(local) observable $A$ and one energy density $e$.

Next we recall some elementary notions from the textbook 
treatment of 
many-body systems in terms of the microcanonical 
and the canonical formalism.
As usual in this context, we thus implicitly take it for granted
that these notions are well-defined and 
that some of their basic qualitative properties 
(see below) are as commonly expected and indeed
observed in many examples.

To begin with, the number of energy eigenvalues below any given 
upper limit $E$ is  
denoted as $\Omega (E) := \sum_n \Theta(E - E_n)$,
where $\Theta(x):=1$ if $x\geq 0$ and $\Theta(x):=0$ otherwise.
More precisely speaking, the step of $\Theta(x)$ at $x=0$
is considered to actually be slightly ``washed out'' such that
$\Omega (E)$ becomes a reasonably smooth function of 
$E$ with a well-defined derivative 
\begin{eqnarray}
\omega(E):=\Omega'(E)=\sum_n\delta(E-E_n)
\ ,
\label{36}
\end{eqnarray}
where also $\delta(x):=\Theta'(x)$ now exhibits
a slightly washed out $\delta$ peak at $x=0$.
Put differently, those washed out $\Theta$ and $\delta$ functions
are understood to be chosen such that $\omega(E)$ can be 
considered as the (local) density of the energy levels $E_n$
(analogously to the discussion 
in the second paragraph of this section).
Once this somewhat subtle point of the standard 
microcanonical formalism is settled, 
the Boltzmann entropy can be defined 
in different (practically) equivalent ways, for instance as 
\begin{eqnarray}
{\cal S}_{\rm B}(E) := k_{\rm B} \ln \Omega(E)
\label{37}
\end{eqnarray}
with Boltzmann constant $k_{\rm B}$.
Finally, the temperature is defined as 
\begin{eqnarray}
T(E) := 1/{\cal S}_{\rm B}'(E)
\, .
\label{38}
\end{eqnarray}

Next, we return to the canonical formalism as 
adopted already in Sec.~\ref{s5}.
In particular, we focus again on the standard canonical 
level populations $p_n$ from Eq.~\eqref{eq:rhoCan}.
Accordingly,
the density operator $\rho_{\rm eq}$ in Eq.~(\ref{16})
can be identified with the canonical ensemble (Gibbs state),
\begin{eqnarray}
\rho_{\rm eq} = Z^{-1} e^{-\beta H}
\, .
\label{42}
\end{eqnarray}
Likewise, $A_{\rm eq}$ in Eq.~(\ref{17}) amounts to the 
corresponding canonical expectation value of $A$,
and is now implicitly understood to be a function of $\beta$.
To make further progress,  
we exploit
that generically the energy levels $E_n$
become extremely (exponentially) dense for large $L$
such that the level populations $p_n$ in Eq.~(\ref{eq:rhoCan})
are practically constant for a large number of closely adjacent 
$E_n$'s. As far as the contributions
to the sum in (\ref{17}) are concerned, one thus can 
approximate the 
large number of corresponding
$A_{nn}$'s by their local average $A_{\rm th}(e_n)$
(see in and below Eq.~(\ref{33})),
yielding
the asymptotically exact large-$L$ approximation
\begin{eqnarray}
A_{\rm eq}(\beta)=\sum_n p_n 
A_{\rm th}(e_n)
\ .
\label{43}
\end{eqnarray}

To make further progress, let us first consider,
as a simple preliminary exercise,
the partition function $Z$, which is defined 
[see also below~\eqref{eq:rhoCan}] for any given 
$\beta>0$ as 
\begin{eqnarray}
Z := \sum_n e^{-\beta E_n}
\ ,
\label{39}
\end{eqnarray}
where we omitted the dependence of $Z$ on $\beta$.
Multiplying the right-hand side by $1=\int dE \, \delta(E-E_n)$ 
and exploiting Eq.~(\ref{36}), one recovers 
the well-known relation
\begin{eqnarray}
Z = \int dE\, \omega(E)\, e^{-\beta E}
\ .
\label{40}
\end{eqnarray}
As usual, in case we are dealing with a model 
exhibiting a finite number $N$ of energy levels
(see above Eq.~({24})), also $\beta\leq 0$
is admissible in Eqs.~(\ref{39}) and (\ref{40}). 
Moreover, the case $\beta<0$ can be readily reduced to the
case $\beta>0$ by considering the Hamiltonian
$-H$ instead of $H$.
Finally, the case $\beta=0$ (infinite temperature) 
is covered by the considerations around Eq.~(\ref{34}).
For simplicity, we therefore continue to focus on 
$\beta>0$.

Returning to the actual quantity of interest in Eq.~(\ref{43}),
we can rewrite it with the help of Eqs.~(\ref{eq:rhoCan}) and (\ref{33}) 
by means of a similar line of reasoning as in the previous paragraph as
\begin{eqnarray}
A_{\rm eq}(\beta)=\int dE A_{\rm th}((E-{\cal E})/L)\, Z^{-1} e^{-\beta E}\omega(E)
\ . \ \ 
\label{44}
\end{eqnarray}
From 
Eqs.~(\ref{36})-(\ref{38})
one furthermore 
can infer that 
$\omega(E)=\Omega(E)/{k_{\rm B}T(E)}$ and thus
\begin{eqnarray}
A_{\rm eq}(\beta) & = & \int dE \, A_{\rm th}((E-{\cal E})/L) \, Q(E)
\ ,
\label{45}
\\
Q(E) & := & [Z\,k_{\rm B}T(E)]^{-1} e^{q(E)}
\ ,
\label{46}
\\
q(E) & := & k_{\rm B}^{-1}{\cal S}_{\rm B}(E) - \beta E
\ ,
\label{47}
\end{eqnarray}
where we omitted the dependence of $Q(E)$ and $q(E)$ on $\beta$.
Going over from the integration variable $E$ to $e:=(E-{\cal E})/L)$ 
(cf. Eq.~(\ref{32})),
and employing the correspondingly transformed quantities
\begin{eqnarray}
\tilde {\cal S}_{\rm B}(e)
& := &
{\cal S}_{\rm B}(eL+{\cal E})/ k_{\rm B} L
\ ,
\label{48}
\\
\tilde T(e)
& := &
k_{\rm B} T(eL+{\cal E})
\ ,
\label{49}
\\
\tilde Q(e)
& := &
L [Z\,
\tilde T(e)]^{-1} e^{\tilde q(e)}
\ ,
\label{50}
\\
\tilde q(e)
& := &
L[\tilde {\cal S}_{\rm B}(e) - \beta e]+\beta {\cal E}
\ ,
\label{51}
\end{eqnarray}
it follows that
\begin{eqnarray}
A_{\rm eq}(\beta) & = & \int de \, A_{\rm th}(e) \, \tilde Q(e)
\ .
\label{52}
\end{eqnarray}
From Eqs.~(\ref{38}), (\ref{48}), and (\ref{49}),
we furthermore can infer 
that $\tilde {\cal S}'_{\rm B}(e) = 1/
\tilde T(e)$.
Hence, the maximum of $\tilde q(e)$ is determined by $\tilde q'(e)=0$,
implying for the maximizing value of $e$, which we denote by 
$e(\beta)$, 
the implicit definition
$\tilde T(e(\beta))=1/\beta$.
Finally, one finds for the second derivative 
(and omitting all arguments $e(\beta)$) that
$\tilde q''=-L \tilde T'/
\tilde T^2=-L\beta^2
\tilde T'$.
Taking for granted the common assumption that the entropy 
density in Eq.~(\ref{48}) exhibits a well-defined thermodynamic 
limit (see also the first two paragraphs of this section),
and that $\tilde T'>0$ (positive heat capacity),
it follows that $\tilde q(e)$ develops a very sharp maximum at
$e(\beta)$ for large $L$.
[Note that this argument would break down for $\beta=0$.]
Finally, choosing $A=\id$ in Eq.~(\ref{52})
implies $\int de\, \tilde Q(e)=1$.
Altogether, $\tilde Q(e)$ 
in Eq.~(\ref{50})
thus approaches a Dirac $\delta$-function
for $L\to\infty$ 
and Eq.~(\ref{52}) yields the asymptotically 
exact large-$L$ approximation
\begin{eqnarray}
A_{\rm eq}(\beta) & = & A_{\rm th}(e(\beta))
\ .
\label{53}
\end{eqnarray}
This finding is very closely related to
the commonly assumed equivalence of the 
canonical and microcanonical ensembles,
see also the discussion
below Eq.~(\ref{eq:rhoCan}).

Analogously, when choosing the canonical level populations (\ref{eq:rhoCan}),
the ETH quantifier in Eq.~(\ref{18}) becomes a function of
$\beta$ and can be shown to satisfy the asymptotically exact 
large-$L$ approximation
\begin{eqnarray}
\Delta^2_{A,{\rm eq}}(\beta) = \Delta_A^2(e(\beta)) 
\ ,
\label{54}
\end{eqnarray}
where $\Delta_A^2(e)$ is defined in Eq.~(\ref{35}).

This is the main result of our present section:
If the canonical ETH quantifier 
$\Delta^2_{A,{\rm eq}}(\beta)$
approaches
a nonzero large-$L$ limit for some $\beta>0$,
then the same applies to 
$\Delta_A^2(e(\beta))$ 
for the specific local observable $A$ under 
consideration and for the specific energy 
density $e(\beta)$ 
(corresponding to the temperature $T=1/k_{\rm B}\beta$).
As explained below Eq.~(\ref{35}),
this implies a violation of the weak ETH.
Regarding the cases $\beta\leq 0$, we
also recall the discussion below Eq.~(\ref{40}).
All of these considerations apply again, in particular, to the example from Sec.~\ref{s5}.

\section{Further Implications}
\label{s6}

\subsection{Equilibration}
\label{s61}

Let us 
consider an arbitrary but fixed initial condition $\rho(0)$, 
which generally may be a pure or mixed 
state far from equilibrium, and which then evolves in time
according to $\rho(t)=e^{-iHt}\rho(0)e^{iHt}$ (in units with $\hbar=1)$.
The expectation value of an observable $A$ at time $t$ is thus given
by $\langle A\rangle_t := \tr\{\rho(t)A\}$ and can be rewritten as
\begin{eqnarray}
	\langle A\rangle_t=\sum_{mn} \rho_{mn}(0) A_{nm} e^{i(E_n-E_m)t}
	\ ,
	\label{19}
\end{eqnarray}
where $\rho_{mn}(t) := \langle m|\rho(t)|n\rangle$ and $A_{nm}:=\langle n|A|m\rangle$.
Systems with many degrees of freedom are
commonly said to exhibit equilibration if the expectation values (\ref{19})
stay close to some constant value for the vast majority of all sufficiently 
late times $t$.
(The excluded minority of times comprises the initial 
relaxation process as well as the very rare but unavoidable 
quantum recurrences.)
This constant value can be identified with the long-time average
of (\ref{19}), which in turn can be rewritten as 
\begin{eqnarray}
	\overline{\langle A\rangle_t}=\sum_{E_m=E_n} \rho_{mn}(0) A_{nm} 
	\ ,
	\label{20}
\end{eqnarray}
where $\overline{f(t)}$ denotes the long-time average 
$\lim_{\tau\to\infty}\frac{1}{\tau}\int_0^t dt\, f(t)$
of an arbitrary function $f(t)$,
and the sum runs over all indices $m$ and $n$ with the property 
$E_m=E_n$.
In the absence of degeneracies, the basis is unique 
(up to trivial phase factors), and the condition
$E_m=E_n$ in (\ref{20})
is tantamount to $m=n$.
However, when $H$ exhibits degeneracies, there are 
many possible choices of the basis and things are 
more involved.
Admitting general initial conditions
$\rho(0)$ (without any further assumptions or restrictions concerning
the matrix elements $\rho_{mn}(0)$ for $m \neq n$),
it follows that the long-time average simplifies to
\begin{eqnarray}
	\overline{\langle A\rangle_t}=\sum_{n} \rho_{nn}(0) A_{nn}
	\label{21}
\end{eqnarray}
if and only if the basis $|n\rangle$ is chosen so that $A$ is diagonal in 
every eigenspace of $H$.
Finally, the result in Eq.~(\ref{21}) may also be rewritten as
$\tr\{\rho_{\rm eq} A\}$, where $\rho_{\rm eq}$ is the equilibrium 
or diagonal ensemble from (\ref{16}) with $p_n:=\rho_{nn}(0)$.

In conclusion, our main finding 
regarding the issue of equilibration is that,
in order to represent the long-time  average in terms of 
a diagonal ensemble,
the pertinent eigenbasis 
of $H$ must maximize the ETH quantifier from (\ref{2}),
meaning that its value is given by $\hat\Delta_A^2$ from~\eqref{17b}.
As an aside it follows that
the corresponding basis $|n\rangle$ and thus the 
diagonal ensemble generally depend on the choice 
of the observable $A$.

\subsection{Autocorrelation functions}
\label{s62}

Next, let us consider a system at equilibrium, described by 
some suitable equilibrium ensemble of the general form (\ref{16}).
Similarly as
the corresponding equilibrium expectation values 
in Eq.~(\ref{17}),
the so-called dynamical auto-correlation function
is defined as
\begin{eqnarray}
	C_{\! A}(t):=\tr\{\rho_{\rm eq}A A(t)\} - A_{\rm eq}^2
	\ ,
	\label{22}
\end{eqnarray}
where $A(t):=e^{iHt}Ae^{-iHt}$ is the 
observable at time $t$ in the Heisenberg picture.
Similarly as in Eqs.~(\ref{19}) and (\ref{20}) it follows that
\begin{eqnarray}
	\overline{C_{\! A}(t)}:=\sum_{E_m=E_n} p_n |A_{nm}|^2 - A_{\rm eq}^2
	\ ,
	\label{23}
\end{eqnarray}
and similarly as in Eq.~(\ref{21}) that 
\begin{eqnarray}
	\overline{C_{\! A}(t)} = \sum_{n} p_n A_{nn}^2 - A_{\rm eq}^2
	\label{24}
\end{eqnarray}
if and only if the basis $|n\rangle$ is chosen such that $A$ 
is diagonal in every eigenspace of $H$.
For this choice,
comparison with Eq.~(\ref{18})
and 
a straightforward calculation
(similarly as in Eq.~(\ref{8})) 
finally yield
\begin{eqnarray}
	\overline{C_{\! A}(t)} = \Delta^2_{A,{\rm eq}} = \hat\Delta^2_{A, {\rm eq}}
	\ ,
	\label{25}
\end{eqnarray}
where we exploited (\ref{18b}) and $p_n=f(E_n)$ (see below Eq.~(\ref{16}))
in the second identity.

In conclusion,
our maximized ETH quantifier
is of immediate relevance for the long-time 
average of the dynamical auto-correlation function.
Notably, this representation of
$\hat\Delta^2_{A, \eq}$ can be used to calculate the basis-independent maximal violation numerically,
	which we
exploited in Fig.~\ref{fig:ETHVio} to evaluate said quantity via dynamical typicality methods (cf.\ Appendix~\ref{app:DynTyp}).

On the other hand,
the time-averaged autocorrelation function $\overline{C_{\! A}(t)}$ can be bounded from below via Mazur inequalities \cite{maz69, mie14, dha21, mou24} in terms of overlaps between the observable $A$ and conserved quantities in the system.
Through the relation~\eqref{25},
the Mazur bounds thus also yield lower bounds for the maximal ETH violation,
providing a connection between the absence of eigenstate thermalization and the 
presence of conservation laws.

\subsection{Rethermalization}
\label{s63}

We focus once more on the canonical ensemble from
Eqs.~(\ref{eq:rhoCan}) or (\ref{42}),
and on the concomitant time-averaged auto-correlation functions as discussed 
in Sec.~\ref{s62}.
Our goal is to show
that these quantities exhibit an interesting connection to the
issue of 
thermalization.

Let us therefore consider a 
modification (perturbation) of our original Hamiltonian $H$ of the form
\begin{eqnarray}
H_{\! g}:=H-gA
\ ,
\label{56}
\end{eqnarray}
where $A$ is a local operator (perturbation) 
and $g$ a small parameter.
[Later this perturbation $A$ will again be identified with the 
observable of interest. Moreover, the coupling parameter $g$ has nothing 
to do with the multiplicities $g_E$ in Sec.~\ref{s31}.]
Similarly as above Eq.~(\ref{19}), the system's initial condition is denoted 
as $\rho(0)$, and we assume that it is given by the canonical ensemble
pertaining to the modified Hamiltonian (\ref{56}), i.e.,
\begin{eqnarray}
\rho(0):=Z_{\! g}^{-1} e^{-\beta H_{\! g}}
\label{57}
\end{eqnarray}
with $Z_{\! g}:=\tr\{e^{-\beta H_{\! g}}\}$.
Physically, the system may be viewed to be 
in thermal equilibrium with respect to the Hamiltonian $H_g$ 
for all times $t<0$, to experience a  
sudden change of the Hamiltonian
at $t=0$ (quantum quench), 
and to be governed by $H$ for $t>0$.
Since $A$ in Eq.~(\ref{56}) is a local operator, 
we are moreover dealing with a so-called 
local quench.
Generally, $\rho(0)$ is thus a non-equilibrium initial state
with respect to the
Hamiltonian $H$ of our actual 
system of interest,
and its time evolution 
is expected to 
exhibit some non-trivial
equilibration process as discussed in the first 
part of Sec. \ref{s61}.

Next we restrict ourselves to the situation that we consider the same
local observable $A$ that also plays the role of the 
perturbation operator in (\ref{56}).
In this 
case
it becomes possible to utilize the generalization 
of Onsager's regression hypothesis from Ref.~\cite{rei24},
consisting in the analytical approximation
\begin{eqnarray}
 \langle A\rangle_t -
  A_{\eq}
 & = &  
g \beta  
\sum_{k=0}^\infty  \frac{(i \beta)^k}{(k+1)!}\,  
\frac{d^k}{dt^k} C_{\! A}(t)
\ .
\label{58}
\end{eqnarray}
Basically, its relative error 
can be shown to become arbitrarily small
for sufficiently small values of $g\beta \opnorm{A}$,
where  $\opnorm{A}$ is the operator norm of $A$ 
(largest eigenvalue in modulus).
A  more detailed and rigorous discussion of this 
error bound
can be found in Refs.~\cite{rei24,rei25u}.

\begin{figure}
\includegraphics[scale=1]{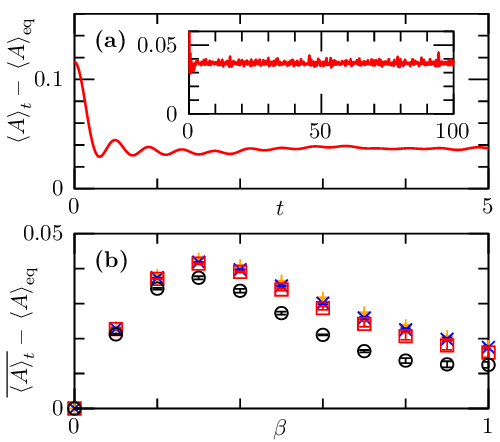}
\caption{(a): Numerically obtained time-dependent expectation values 
$\langle A\rangle_t -  A_{\eq}$ for the observable $A = s_1^z$, 
the Hamiltonian~\eqref{55}, the initial condition (\ref{57}),
and the canonical ensemble~\eqref{42}
with parameter values $L=15$, $g=1$, and $\beta=0.2$,
see also Eqs.~(\ref{17}), (\ref{19}), and (\ref{56}).
Main plot: Initial relaxation behavior for $t\in[0,5]$.
Inset: Long-time behavior for $t\in[0,100]$.
(b): Long-time average of the same expectation values 
as in (a) versus $\beta$.
Orange crosses, blue crosses, and red squares 
correspond to $L=11$, $13$, and $15$, respectively.
Black circles with error bars: Extrapolations for
$L\to\infty$.
For further details regarding the numerical methods
and the extrapolations we refer to Fig.~\ref{fig:ETHVio}
and its discussion in the main text.}
\label{fig2}
\end{figure}

Of particular interest in our present context is the long-time average,
see in and around Eq.~(\ref{20}).
Under very weak assumptions regarding the long-time behavior of 
$C_{\! A}(t)$, which we tacitly take for granted, 
the long-time average of 
$d^k C_{A}(t)/d^k t$ vanishes for all $k\geq 1$, and
we can conclude from Eq.~(\ref{58}) that
\begin{eqnarray}
\overline{\langle A\rangle_t} -
A_{\eq}
& = &
g\beta \,\overline{C_{\! A}(t)}
\ .
\label{59}
\end{eqnarray}
The difference on the left-hand side quantifies how well the
expectation values $\langle A\rangle_t$ approach the
canonical equilibrium value 
$A_{\eq}$
in the long run (see also the discussion 
below Eqs.~(\ref{42}) and (\ref{19})).
In particular, if this difference turns out to remain finite
in the thermodynamic limit, we can conclude that the
system does not exhibit thermalization \cite{note:EquilibriumState}.
According to Eq.~(\ref{25}) and the results from
Sec.~\ref{s5}, it follows that this is indeed the case
for the particular Hamiltonian from Eq.~(\ref{55}).
In the same way, absence of thermalization can be
deduced in full generality whenever
the considered model violates the weak ETH
in some basis.

In conclusion,
the existence of a basis in which the weak ETH is satisfied,
as demonstrated in Refs.~\cite{bir10, mor16, iyo17, mor18, kuw20, kuw20a} 
for translationally invariant systems,
does not guarantee rethermalization after a local quench
for systems with degeneracies,
even though it is sufficient for systems without degeneracies \cite{far17,dab22,pil24}.
Rethermalization can only be established if the basis-independent upper bound $\hat\Delta^2_{A, \eq}$ from~\eqref{18b} vanishes in the thermodynamic limit.
The physically relevant generalization of the weak ETH should thus be formulated in terms of $\hat\Delta^2_{A, \eq}$ or $\hat\Delta^2_A$ from~\eqref{17b}.
Unfortunately, we are not aware of any proofs of such $\hat\Delta_A^2 \to 0$ statements for general classes of systems.

A numerical illustration of these general analytical predictions
is provided in Fig.~\ref{fig2} for the same model 
system as in Fig.~\ref{fig:ETHVio}.
Fig.~\ref{fig2}(a) depicts the initial temporal relaxation process
and the approach of an approximately steady long-time value
which is clearly different from the thermal equilibrium value.
The remnant small fluctuations (inset) are a well-known 
finite-size effect.
Fig.~\ref{fig2}(b) shows the corresponding long-time averages 
for various different $\beta$ and $L$ values.
Once again this exemplifies the analytically predicted
absence of rethermalization after a local quench 
(except for the trivial case $\beta=0$).
Moreover, upon comparison with the results 
from Fig.~\ref{fig:ETHVio} one 
could also confirm that
the analytical approximation (\ref{59}) (see also Eq.~(\ref{25}))
is indeed quite well satisfied. 
We omit here such a quantitative comparison since it goes beyond 
the actual main themes of our present work.

\subsection{Degeneracies and numerical ETH studies}
\label{s64}

\begin{figure}
\includegraphics[scale=1]{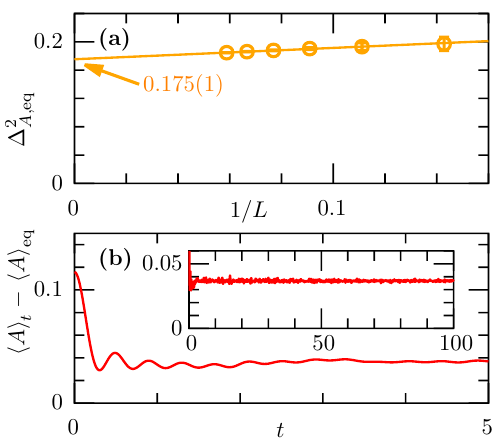}
\caption{(a) Same as the middle panel ($\beta = 0.2$) of Fig.~\ref{fig:ETHVio}, but now for the perturbed, nondegenerate 
Hamiltonian $H'$ from Eq.~\eqref{eq:H'}.
(b) Same as Fig.~\ref{fig2}(a), but 
now again
for the 
perturbed system $H'$.}
\label{fig:perturbations}
\end{figure}

The basis dependence of eigenstate thermalization observed 
here can only occur in many-body systems with degeneracies,
which typically 
have their origin in some symmetries 
(or corresponding conserved quantities).
As usual in the context of symmetry considerations,
their implications are generally regarded as very interesting and useful
in spite of the fact that
these
symmetries 
may only be 
approximately
satisfied in realistic systems.
This means
that degeneracies are also only approximate,
yet they
can still have significant,
macroscopically observable effects on the system.
For example, if a system with exact symmetries fails to thermalize,
a nearby, approximately symmetric cousin could
still fail to thermalize, or it could exhibit extremely slow relaxation 
and fail to thermalize on any practically relevant time scale \cite{ber04, kin06, moe08, eck09, ber11, gri12, mal19, rei19}.

The critical question in 
our present case 
is whether the ETH quantifiers~\eqref{2} and~\eqref{18} in a weakly perturbed 
system without degeneracies approach zero or a nonzero value in the thermodynamic limit.
In the former case, choosing the translationally invariant basis 
in the idealized (unperturbed) system
would still yield the right conclusions for 
the more realistic (slightly perturbed)
systems without degeneracies.
In the latter case, however,
any conclusions drawn from the ETH 
quantifiers
in the translationally 
invariant eigenbasis would be wrong not only in the 
idealized (translationally invariant) system itself,
but also in the
weakly perturbed 
modifications
without degeneracies.

To test this,
we perturb the Hamiltonian~\eqref{55} by adding local magnetic fields in the $z$ direction with amplitudes $h_l$ drawn from a normal distribution with vanishing mean and standard deviation $0.1$,
\begin{equation}
\label{eq:H'}
	H' = - \sum_{l=1}^L \left[ s_l^+ (s_{l+1}^-)^2 s_{l+2}^+ + s_l^- (s_{l+1}^+)^2 s_{l+2}^- + h_l \, s_l^z \right] .
\end{equation}
The additional fields break the translation and spatial reflection symmetries,
and we have verified numerically that this system indeed exhibits no degeneracies.

The results in Fig.~\ref{fig:perturbations}(a) demonstrate that the 
violation of the weak ETH 
persists in the nondegenerate regime 
since the quantity
$\Delta_{A,\eq}^2$ from~\eqref{18}, now being unique, approaches
$0.175$
as $L \to \infty$.
Consequently, the system cannot be expected to rethermalize after a local quench,
and Fig.~\ref{fig:perturbations}(b) indeed confirms this absence of thermalization explicitly for a similar quench setup as in Fig.~\ref{fig2}.

The crucial consequence of these results is that exploiting symmetries to evaluate
the weak ETH numerically in a system with degeneracies can lead to physically wrong conclusions:
Using translationally invariant eigenstates to calculate $\Delta_{A,\eq}^2$ 
from~\eqref{18} in the system from~\eqref{55} would suggest that the weak ETH is satisfied,
and one could be tempted to conclude that this system rethermalizes after local quenches.
This conclusion would be wrong for both the symmetric, idealized case with degeneracies and the close-by, weakly perturbed versions without degeneracies from~\eqref{eq:H'}.

In principle, analogous problems may arise also for the strong
(or any other) version of the ETH
and numerical studies thereof.
For example, if one evaluates the diagonal matrix elements of some observable
by exploiting translational invariance in a system that is also reflection symmetric 
(and thus exhibits energy degeneracies, cf.\ Sec.~\ref{s2}),
there is no guarantee that 
a so-observed 
agreement with the strong ETH 
is inherited by
the physically relevant basis in which the observable is diagonal in degenerate subspaces.
Likewise, there is no guarantee that physically realistic, weakly perturbed, nondegenerate variations 
of such a system resemble the unperturbed behavior
observed in 
the translation-symmetric basis.

\subsection{Conserved quantities and symmetry sectors}
\label{s66}

\begin{figure}
	\includegraphics[scale=1]{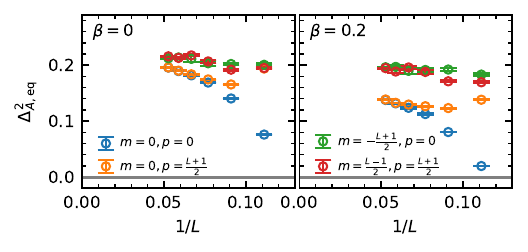}
	\caption{
		Same as in Fig.~\ref{fig:ETHVio} but now employing the maximally ETH-violating basis within various magnetization ($m$) and dipole-moment ($p$) subsectors (see also Appendix~\ref{app:DynTyp}).
	}
	\label{fig:ETHVioSubsectors}
\end{figure}

It is widely 
taken for granted
that information about conserved quantities represented by local extensive observables (sums of local operators) needs to be included to describe a system's long-term properties.
The thermal state for a system conserving the total magnetization or the number of particles, for example,
accounts for this 
either via appropriate restrictions
to a macroscopically small window of states with similar expectation values of the conserved quantities (microcanonical approach),
or via
additional Lagrange multipliers besides the inverse temperature (grandcanonical approach) \cite{lan70}.
Likewise, equilibrium properties of integrable systems are described by generalized Gibbs ensembles (GGEs) that include their extensive number of conserved quantities \cite{rig07, lan15, vid16}.

Remarkably, previous results in the context of weak eigenstate thermalization (e.g., the proofs from Refs.~\cite{bir10, mor16, iyo17, mor18, kuw20, kuw20a} for translationally invariant eigenbases)
do \emph{not} need to include information about conserved quantities. 
However,
tests of the strong ETH should usually be restricted to symmetry sectors of local conserved quantities \cite{san10, mor18, mon18}.
On the other hand, it is widely believed that conserved quantities represented by nonlocal and/or many-body operators usually do not need to be considered when assessing the diagonal ETH \cite{san10, kim14, mon18, mor18}.
This distinction is crucial because any quantum system exhibits as many conserved quantities as energy levels (take the projectors onto energy eigenspaces),
so without a criterion for what constitutes physically relevant and irrelevant conserved quantities,
a statement like the ETH would become void.

Our example system from Sec.~\ref{s5} conserves the energy and the (nonlocal) quasimomentum and spatial parity.
It also exhibits two additional local conserved quantities, the magnetization and the dipole moment in the $z$ direction (see also below Eq.~\eqref{55} and Appendix~\ref{app:DynTyp}).
These were not taken into account in Fig.~\ref{fig:ETHVio}.
For completeness, we show in Fig.~\ref{fig:ETHVioSubsectors} that the restriction to simultaneous magnetization and dipole-moment subsectors does not remove the ETH violation in the most dangerous basis in this model.
On the other hand,
it is impossible to restrict to a simultaneous subsector of definite magnetization, dipole moment and quasimomentum
because the latter two quantities do not commute.

A series of recent works have suggested extensions of thermal states \cite{yun16, gur16, los17} and the ETH \cite{mur23, pat25, las24} in the presence of noncommuting conserved quantities (see also the review~\cite{maj23}),
which
gives
rise to degeneracies among other subtleties (cf.\ Sec.~\ref{s2}).
However, the mechanisms identified there, and the fixes suggested,
do not apply to our present problem of basis-dependent ETH quantifiers.
In all of these works,
it is implicitly or explicitly assumed that the underlying symmetries are 
continuous and that the resulting conserved quantities are extensive operators.
In particular,
the equilibrium value $\< A \>_{\mathrm{eq}}$ used in the examples 
from Figs.~\ref{fig2} and~\ref{fig:perturbations} 
coincides with the value
that would be obtained from a non-Abelian thermal 
state (NATS) \cite{mur23, pat25, las24} constructed with the (nonlocal) 
conserved quasimomentum taken into account.
The reason is that the
expectation value of this operator vanishes in the 
initial state~\eqref{57},
hence the associated Lagrange multiplier is zero in the NATS, and thus 
it coincides with the equilibrium state from~\eqref{16} and~\eqref{eq:rhoCan}.
In other words, the NATS concept does not help to recover the correct
long-term limit of the dynamics in Figs.~\ref{fig2} 
and~\ref{fig:perturbations}.

Specifically in the context of non-Abelian ETH generalizations, the focus has mostly been on $\mathrm{SU}(2)$ symmetry in spin-$\frac{1}{2}$ models.
For such a Heisenberg-type model, it was numerically observed in Ref.~\cite{pat25} 
that the generalized ETH for noncommuting charges can be evaluated across 
quasimomentum/translational-symmetry sectors,
just like in the traditional ETH framework.

For the possibility of basis-dependent ETH quantifiers
-- 
the key observation of our present paper
--
the origin of the degeneracies does not matter.
As we demonstrated in Sec.~\ref{s2},
excessive degeneracies can 
already result 
solely
from the 
interplay of symmetries with nonlocal conserved quantities (quasimomentum and spatial parity),
which previous studies usually disregarded as irrelevant for (diagonal) ETH 
explorations
\cite{san10, kim14, mon18, mor18, pat25}.

In principle, one still
might hope that this 
theoretically possible
basis dependence of eigenstate thermalization 
is irrelevant in practice because the scaling of quantities like $\Delta_A^2$ 
from~\eqref{2} and $\Delta_{A,\eq}^2$ from~\eqref{18} with the system size could be identical for all basis choices.
We demonstrate that this hope is in vain with our example in Fig.~\ref{fig:ETHVio},
and that this has practical implications as expounded in the previous subsections and 
Figs.~\ref{fig2} and~\ref{fig:perturbations} in particular.

In summary, while our 
present explorations are to some extent related
to previously reported examples and reasons for
ETH violations,
our
main point is actually a different one:
The very question of whether a system satisfies or violates the ETH is 
generally ambiguous and thus physically ill-posed in systems with 
degeneracies.

\section{Summary and Conclusions}
\label{s7}

Various 
versions of the ETH have been 
proposed, and some of them are widely considered to 
explain why many-body quantum systems generically thermalize.
Essentially, they consist in 
predictions regarding the diagonal matrix elements of local observables 
in the eigenbasis of the system's Hamiltonian.
In general, the choice of the eigenbasis, and hence the properties of these diagonal
matrix elements, are not unique, while physical system properties cannot depend on 
that choice.
Given thermalization is such a property, and that some 
version of the 
ETH is supposed to explain it, also the answer to the question of whether 
the ETH is fulfilled or violated must be basis independent.

The upshot of our paper is that whether or not a given many-body
Hamiltonian
satisfies some form of ETH may in general 
depend on the choice of the energy eigenbasis.
Specifically, we presented an explicit example where the weak ETH 
is satisfied in one basis and violated in another basis.
Put differently (see Sec.~\ref{s1}), eigenstate thermalization
is violated by a 
nonzero
fraction of all eigenstates
in one basis and by a vanishingly small fraction in 
the other basis.
These findings explicitly demonstrate that suitable, 
basis-independent generalizations of the ETH are needed 
to retain its 
presumed
role as the fundamental mechanism behind thermalization.

An obvious precondition for
basis-dependent ETH quantifiers
is that the considered 
Hamiltonian $H$ exhibits degeneracies. 
We  
showed
that the vast majority of all eigenvalues
indeed exhibit degeneracies if $H$ is translationally 
symmetric as well as invariant under a spatial 
reflection.
The relevance of this result becomes apparent by
observing that translationally invariant models with 
broken reflection symmetry are actually 
very uncommon.
Yet, they do exist, as exemplified by so-called ratchet 
models \cite{rei02}, or the models considered 
in Ref. \cite{tas24}.
An important implication is that previous results which have been obtained
under the assumptions of translational invariance and
absence of degeneracies,
for instance
in Refs.~\cite{far17,dab22,alh20,pil24},
are rigorously speaking 
restricted 
to 
such ``uncommon'' models (without reflection symmetry).
Finally, even in those latter cases there still may
exist some other symmetries from which one may again infer the
existence of degeneracies along similar lines as 
in Sec.~\ref{s2}.

In a next step, we
demonstrated that the ETH quantifier from
Eq.~(\ref{2}), or more generally from Eq.~(\ref{18}),
assumes its maximally possible value if the basis
is chosen such that the considered observable $A$ is diagonal 
in every eigenspace of $H$.
In turn,  the ETH quantifier assumes its minimally possible 
value when choosing the basis such that the diagonal matrix 
elements of $A$ are constant within every eigenspace of $H$.

The 
maximizing bases were 
also
found to be
of considerable importance in several other contexts: 
They must be chosen in order to express 
long-time averages in terms of a so-called diagonal 
ensemble, see Eq.~(\ref{21}).
Likewise, they govern the long-time average of
dynamical auto-correlation functions, see 
Eq.~(\ref{25}).
The latter equation was demonstrated 
in Appendix \ref{app:DynTyp} to be
also very useful for numerical 
explorations of the ETH.
Finally, the maximizing bases play a key role regarding 
the question of whether a system rethermalizes after 
a local quench, see Eq.~(\ref{59}).

We furthermore showed that these physical consequences remain observable in weakly perturbed systems without degeneracies.
In particular, one may draw the wrong physical conclusions if one evaluates the ETH by exploiting symmetries in systems with degeneracies.
This insight is especially relevant for numerical studies of the ETH.

Since the values assumed by the ETH quantifiers~\eqref{2} and~\eqref{18} in the maximizing bases can be expressed in terms of basis-independent quantities, cf.\ Eqs.~\eqref{17b} and~\eqref{18b}, 
the latter suggest themselves as natural, unambiguous generalizations of weak ETH quantifiers.
It will be important to characterize these objects in more detail and possibly derive conditions under which their vanishing in the thermodynamic limit can be guaranteed,
extending such proofs for the traditional, basis-dependent quantifier~\eqref{2} for translationally invariant eigenbases.

Specifically, for our above-mentioned example
of a translationally invariant system which 
violates the weak ETH in some energy eigenbasis,
it follows that the auto-correlation
function does not decay to zero for asymptotically large times,
and that the system does not rethermalize after a 
local quench.
Since such behavior has not been commonly observed
until now,
it seems a quite interesting and challenging task 
to discover and explore  larger classes of such 
examples in the future.
Another formidable task may be to find analogous
examples where the conventional (strong) ETH is
fulfilled in one basis and violated in another basis.

\begin{acknowledgments}
This work was supported by the 
Deutsche Forschungsgemeinschaft (DFG, German Research Foundation)
under Grant No. 355031190 
within the Research Unit FOR 2692
and under Grant No. 
502254252.
This research utilized Queen Mary's Apocrita HPC facility, 
supported by QMUL Research-IT \cite{but17}.
This work was also supported by the Paderborn Center for Parallel 
Computing (PC$^2$) within the project 
HPC-PRF-UBI2.
\end{acknowledgments}

\section*{Data availability statement}
Source data for all the figures are available at Ref.~\cite{fig_source_data}.

\vspace*{0.0cm}
\appendix
\section{Derivation of Eq. (\ref{6})}
\label{app1}

In this Appendix, we provide the derivation of Eq.~(\ref{6}).
In doing so, we tacitly employ the same notation and 
definitions as in Sec.~\ref{s2}.

For any given system size $L \geq 2$,
its unique prime factor decomposition 
(PFD) can be written in the form
\begin{eqnarray}
L=\prod_{j=1}^J (p_j)^{\mu_j}
\label{a1}
\end{eqnarray}
for suitably chosen exponents  $\mu_j\in\NN$,  
prime numbers $p_j$, and $J\in\NN$.
Since $p_j\geq 2$ and $\mu_j\geq 1$ it follows that
$(p_j)^{\mu_j}\geq 2$ for all $j\in\{1,...,J\}$.
Together with (\ref{a1}) this yields $L\geq 2^J$
and thus 
\begin{eqnarray}
J\leq \ln L/\ln 2
\ .
\label{a2}
\end{eqnarray}

Recalling the definitions (\ref{3}) and (\ref{4}),
our next goal is to estimate the number of basis
vectors $|{\bm\sigma}\rangle$ with the property that
\begin{eqnarray}
T^l |{\bm\sigma}\rangle = |{\bm\sigma}\rangle
\label{a3}
\end{eqnarray}
for some suitably chosen 
$\l\in\{1,...,L-1\}$.
Therefore, let $|{\bm\sigma}\rangle$ be an arbitrary but 
fixed basis vector for which there exists an 
$\l\in\{1,...,L-1\}$ with the property (\ref{a3}).
Without loss of generality we furthermore assume
that $l$ has been chosen as small as possible.
Since $1\leq l<L$ and $L\geq 2$
one can conclude that there must exist 
an integer $m\geq 2$ with the
property $ml=L+r$ and $0 \leq r <l$.
Recalling $T^L =1$
we thus can infer that
$T^{r} |{\bm\sigma}\rangle 
= T^{r+L} |{\bm\sigma}\rangle 
= T^{ml} |{\bm\sigma}\rangle 
= |{\bm\sigma}\rangle$.
Since $l$ is the smallest integer within
$\{1,...,L-1\}$ with the property (\ref{a3})
we can infer that $r=0$ and thus
$ml=L$. 
Since the PFD of $L$ in (\ref{a1}) is unique, and 
likewise for the PFD of $m\geq 2$ and of $l$,
the relation $ml=L$ implies that
at least one $p_j$ must 
also appear in the PFD of $m$ and hence 
$m':=m/p_j$ must be a positive integer.
It follows 
that $m'l=L/p_j$ must be a 
positive integer
satisfying 
$T^{L/p_j} |{\bm\sigma}\rangle
=T^{m'l} |{\bm\sigma}\rangle 
= |{\bm\sigma}\rangle$.
In other words,
\begin{eqnarray}
q_j:=L/p_j
\label{a4}
\end{eqnarray}
must be an integer with $1\leq q_j \leq L/2$,
where we exploited that $p_j\geq 2$ 
in the last step.
The overall conclusion is that
for every $|{\bm\sigma}\rangle$ 
with the property (\ref{a3}) 
for some $\l\in\{1,...,L-1\}$
there exists at least one $j\in\{1,...,J\}$
with the property
\begin{eqnarray}
T^{q_j} |{\bm\sigma}\rangle = |{\bm\sigma}\rangle
\ .
\label{a5}
\end{eqnarray}

Recalling that there are $D^L$
orthonormalized basis vectors of the form
(\ref{3}),
there must exist exactly $D^{q_j}$ different
basis vectors $|{\bm\sigma}\rangle$ with the property
(\ref{a5}) for the following reason:
For each $l$ with $1\leq l\leq q_j$ 
we can choose $\sigma_l$ arbitrarily in $\{1,...,D\}$, 
while all $\sigma_l$ with $l>q_j$ are
then uniquely fixed via (\ref{a5}).
Denoting by $N_{\rm per}$ the total number of all 
$|{\bm\sigma}\rangle$ satisfying (\ref{a3}) for some $l\in\{1,...,L-1\}$,
we can conclude that $N_{\rm per}\leq
\sum_{j=1}^J D^{q_j}$.
(The inequality is a consequence of the fact that
some $|{\bm\sigma}\rangle$ may satisfy (\ref{a5}) 
simultaneously for several different
$j\in\{1,...,J\}$.)
Since $q_j\leq L/2$ (see below (\ref{a4})) it follows that
$N_{\rm per}\leq 
J D^{L/2}$ and with (\ref{a2}) that
\begin{eqnarray} 
N_{\rm per}\leq 
D^{L/2}\ln L/\ln 2
\ .
\label{a6}
\end{eqnarray}

The above considerations imply that
$\bar N_{\rm per}:=D^L-N_{\rm per}$ 
is the number of all basis vectors 
$|{\bm\sigma}\rangle$ with the property 
\begin{eqnarray}
T^l |{\bm\sigma}\rangle \not = |{\bm\sigma}\rangle
\ \ \mbox{for all $l\in\{1,...,L-1\}$.}
\label{a7}
\end{eqnarray}
Moreover, it follows with 
(\ref{a6})
that
\begin{eqnarray} 
\bar N_{\rm per}
& = & 
(1-\delta)\, D^L
\label{a8}
\end{eqnarray}
for some suitably chosen $\delta$ with
\begin{eqnarray}
\delta
& \leq & \frac{\ln L}{D^{L/2}\ln 2}
\ .
\label{a9}
\end{eqnarray}
Altogether, we thus can conclude that for large $L$ the vast majority of all
the $D^L $ basis vectors $|{\bm\sigma}\rangle$ 
satisfies
(\ref{a7}).

Observing that $\langle {\bm\sigma}'| {\bm\sigma}\rangle$ 
is unity if $|{\bm\sigma}'\rangle=|{\bm\sigma}\rangle$ 
and zero otherwise, 
we can conclude that every $|{\bm\sigma}\rangle$ with the 
property (\ref{a7}) satisfies
$\langle T^m{\bm\sigma}|T^n{\bm\sigma}\rangle=\delta_{mn}$
for all $m,n\in\{0,...,L-1\}$.
In other words, the $L$ vectors $\{T^l |{\bm\sigma}\rangle\}_{l=0}^{L-1}$
are normalized and pairwise orthogonal.

For any given $|{\bm\sigma}\rangle$ with 
the property (\ref{a7}) we denote by $\hr_{\bm\sigma}$ the subspace
which is spanned by the $L$ 
vectors $\{T^l |{\bm\sigma}\rangle\}_{l=0}^{L-1}$.
Since $T^L |{\bm\sigma}\rangle =|{\bm\sigma}\rangle$ it follows that
$\hr_{{\bm\sigma}'}=\hr_{\bm\sigma}$ whenever 
$|{\bm\sigma}'\rangle$ can be 
written as $T^l |{\bm\sigma}\rangle$ for some $l$.
Hence there must be $\bar N_{\rm per}/L$ pairwise orthogonal 
subspaces $\hr_{\bm\sigma}$, and each $|{\bm\sigma}\rangle$ 
with the property (\ref{a7}) is contained in 
exactly one of them.

For an arbitrary but fixed $|{\bm\sigma}\rangle$ with 
the property (\ref{a7}) and any $\nu\in\{0,...,L-1\}$
we define
\begin{eqnarray}
|{\bm\sigma}_\nu\rangle := L^{-1/2} \sum_{l=0}^{L-1} e^{-2\pi i \frac{\nu l}{L}}\, T^l |{\bm\sigma}\rangle
\ .
\label{a10}
\end{eqnarray}
Exploiting 
that the $\{T^l |{\bm\sigma}\rangle\}_{l=0}^{L-1}$
are orthonormalized (see above),
a straightforward calculation yields
\begin{eqnarray}
\langle {\bm\sigma}_\mu |{\bm\sigma}_\nu\rangle = \delta_{\mu\nu}
\label{a11}
\end{eqnarray}
for all $\mu,\nu\in\{0,...,L-1\}$.
Recalling that $\hr_{\bm\sigma}$ is spanned by $\{T^l |{\bm\sigma}\rangle\}_{l=0}^{L-1}$,
we thus can conclude that $\{|{\bm\sigma}_\nu\rangle \}_{\nu=0}^{L-1}$ is 
an
orthonormal basis of $\hr_{\bm\sigma}$.
Moreover, one can infer from (\ref{a10}) and $T^L=1$ that
\begin{eqnarray}
T |{\bm\sigma}_\nu\rangle = \gamma_\nu |{\bm\sigma}_\nu\rangle
\label{a12}
\end{eqnarray}
with $\gamma_\nu$ from Eq.~(\ref{5}).

Finally, $T^L=1$ implies that all eigenvalues of $T$ are of the form
(\ref{5}) with $\nu\in\{0,...,L-1\}$.
For an arbitrary but fixed $\nu\in\{0,...,L-1\}$
we denote by $\hr_\nu$ the eigenspace of $T$
with eigenvalue $\gamma_\nu$.
Since each of the  $\bar N_{\rm per}/L$ pairwise orthogonal 
subspaces $\hr_{\bm\sigma}$ contributes one eigenvector of the
form (\ref{a10}) to $\hr_\nu$ it follows that the dimension of
$\hr_\nu$ is lower bounded by $\bar N_{\rm per}/L$.
Together with (\ref{a9}) and (\ref{a8}) we finally recover
Eq.~(\ref{6}).

\section{Symmetry considerations for the Hamiltonian $H$ from Eq.~(\ref{55})}
\label{app2}

This Appendix collects various symmetry properties of the
specific spin-chain model from Eq.~(\ref{55}) and certain 
generalizations thereof.
In particular, we will identify a large class of observables $A$,
including $A=s_l^z$, for which the thermal expectation value in 
(\ref{1}) vanishes 
for any choice of the microcanonical energy window.
For the rest, we utilize the same notation as in the main 
paper  without recalling all the pertinent definitions.

To make the paper self-contained, we begin with a
re-derivation of the previously known identity
\begin{eqnarray}
e^{i\pi s^x}s^{y,z}=-s^{y,z}e^{i\pi s^x}
\label{b1}
\end{eqnarray}
for spin operators $s^{x,y,z}$ with arbitrary integer or 
half-integer spins in units with $\hbar=1$:

From the definition $e^X:=\sum_{k=0}^\infty X^k/k!$ one can infer the
so-called Hadamard lemma of the Baker-Campbell-Hausdorff formula,
reading
\begin{eqnarray}
e^X Y e^{-X} & = &\sum_{m=0}^\infty \frac{1}{m!}\,[X,Y]_m
\ ,
\label{b2}
\\
\mbox{$[X,Y]_0$} & := & Y
\ ,
\label{b3}
\\
\mbox{$[X,Y]_{m+1}$} & := & [X, [X,Y]_m]
\ ,
\label{b4}
\end{eqnarray}
where $[\,\cdot\,,\cdot\,]$ denotes the usual commutator.
Choosing 
\begin{eqnarray}
X & := & q s^x
\ ,
\label{b5}
\\
Y & := & q s^y
\ ,
\label{b6}
\\
Z & := & q s^z
\ ,
\label{b7}
\end{eqnarray}
with an arbitrary but fixed $q\in\CC$,
and exploiting the usual
commutation relations for $s^{x,y,z}$ it follows that
\begin{eqnarray}
[X,Y]_1 & = & [X,Y] = i q Z
\ ,
\label{b8}
\\
\mbox{$[X, Y]_2$} & = &  [X, iqZ] = q^2 Y = q^2 [X,Y]_0
\ ,
\label{b9}
\\
\mbox{$[X, Y]_3$} & = &  [X, q^2 Y] = i q^3 Z = q^2 [X,Y]_1
\ ,
\label{b10}
\end{eqnarray}
and so on.
Together with (\ref{b3}) and (\ref{b4}) one can conclude that
\begin{eqnarray}
[X,Y]_{m} & = & \mbox{$q^m Y $ for even $m$}
\ ,
\label{b11}
\\
\mbox{$[X, Y]_m$} & = & \mbox{$i q^m Z$ for odd $m$}
\ ,
\label{b12}
\end{eqnarray}
and with (\ref{b2}) that
\begin{eqnarray}
e^X Y e^{-X} & = &Y \cosh(q) + i Z \sinh(q)
\ .
\label{b13}
\end{eqnarray}
Exploiting (\ref{b5})-(\ref{b7}) and setting $q=i\phi$ this yields
\begin{eqnarray}
e^{i\phi s^x} s^y e^{-i\phi s^x} & = & s^y \cos(\phi) - s^z \sin(\phi)
\ .
\label{b14}
\end{eqnarray}
In the special case $\phi=\pi$ we thus recover (\ref{b1}) for $s^y$.
Analogously, one verifies (\ref{b1}) for $s^z$.

For later use we also note the obvious relations
\begin{eqnarray}
e^{i\pi s^x}s^{x}=s^{x}e^{i\pi s^x}
\ ,
\label{b15}
\\
\mbox{$[e^{i\pi s^x}]^\dagger$}= e^{-i\pi s^x} = [e^{i\pi s^x}]^{-1}
\ .
\label{b16}
\end{eqnarray}

Let us now consider any spin Hamiltonian on a lattice $\Lambda$
of the general structure
\begin{eqnarray}
H=\sum_\nu h_\nu
\ ,
\label{b17}
\end{eqnarray}
where each summand $h_\nu$ consists of a product of factors 
of the form $s_l^a$ with indices $l\in\Lambda$ and $a\in\{x,y,z\}$
which may or may not be different for each factor.
Our only requirement is that {\em the number of factors $s_l^a$
with the property $a\in\{y,z\}$ must be even for each $h_\nu$.}
All further details, for instance concerning the lattice $\Lambda$,
the set of indices $\nu$, the boundary conditions, or
translational and parity symmetries,
will play no role in the following.

Our second key player will be the operator
\begin{eqnarray}
U:=\exp\{i\pi\sum_{l\in\Lambda} s_l^x\}
\ .
\label{b18}
\end{eqnarray}
Since all $s_l^x$ commute with each other, one readily infers from
(\ref{b1}), (\ref{b15}), (\ref{b16}) that
\begin{eqnarray}
U s_l^x & = & s_l^x U
\ ,
\label{b19}
\\
U s_l^y & = & - s_l^y U
\ ,
\label{b20}
\\ 
U s_l^z & = & - s_l^z U
\ ,
\label{b21}
\\
U^\dagger & = & U^{-1}
\label{b22}
\end{eqnarray}
for any $l\in\Lambda$.
Since we assumed that the number of factors $s_l^a$ with the 
property $a\in\{y,z\}$ is even for each $h_\nu$,
one can infer that $Uh_\nu=h_\nu U$ and with (\ref{b17}) that
\begin{eqnarray}
UH=HU
\ .
\label{b23}
\end{eqnarray}

Let $|n\rangle$ be an arbitrary but fixed eigenvector of $H$ with eigenvalue $E_n$
and define $|\tilde n\rangle:=U|n\rangle$.
It readily follows with (\ref{b23}) that $H|\tilde n\rangle = E_n|\tilde n\rangle$
and with (\ref{b21}), (\ref{b22}) that 
\begin{eqnarray}
\langle \tilde n| s_l^z |\tilde n\rangle= - \langle n| s_l^z  |n\rangle
\label{b24}
\end{eqnarray}
for any $l\in\Lambda$.
In conclusion, either $\langle n| s_l^z  |n\rangle=0$ or there must be a pair
of degenerate eigenvectors with opposite diagonal elements of the 
observable $s_l^z$.

The remaining problem is that $\langle n|\tilde n\rangle$ may not be zero, 
meaning that $|n\rangle$ and $|\tilde n\rangle$ may not be members of the 
same orthonormal basis (ONB).
The way out is as follows: Let us denote by $\hr_{\! E}$ an arbitrary but fixed 
eigenspace of $H$ and by $E$ the corresponding eigenvalue.
In the rest of this paragraph we temporarily restrict ourselves to 
eigenvectors $|n \rangle$ belonging to this
eigenspace $\hr_{\! E}$.
For every such $|n\rangle\in\hr_{\! E}$ we define its counterpart
$|\tilde n\rangle:=U|n\rangle$.
As before, it follows  that all those $|\tilde n\rangle$ are 
again eigenvectors of $H$ with eigenvalue $E$ and thus are contained 
in $\hr_{\! E}$. Moreover, one finds that $\langle n| m\rangle=\langle \tilde n|\tilde m\rangle$,
i.e., both the $|n\rangle$ and the $|\tilde n\rangle$ are an ONB of $\hr_{\! E}$. 
As usual, the trace over the (sub-)space $\hr_{\! E}$ is basis independent, i.e.,
\begin{eqnarray}
\tr_{\hr_{\! E}} \{s_l^z\}\ 
= \sum_n \langle n| s_l^z | n\rangle 
=  \sum_{n} \langle \tilde n| s_l^z |\tilde n\rangle 
\ .
\label{b25}
\end{eqnarray}
Together with (\ref{b24}) it follows that
\begin{eqnarray}
\tr_{\hr_{\! E}} \{s_l^z\} = \sum_n \langle n| s_l^z | n\rangle = 0
\label{b26}
\end{eqnarray}
for any $l\in\Lambda$ and for any eigenbasis $|n\rangle$ of $\hr_{\! E}$.

Returning to the full Hilbert space $\hr$ of $H$ and the corresponding
full ONB $|n\rangle$, we can conclude that the average of the 
diagonal elements  $\langle n| s_l^z | n\rangle$ 
over any (arbitrarily small or large) energy interval 
is always zero.
In other words, we recover the announced property
that the thermal expectation value of $A=s_l^z$ in 
(\ref{1}) vanishes for any choice of the microcanonical 
energy window.
Likewise, one finds that the equilibrium expectation value $A_{\rm eq}$
from Eq.~(\ref{17}) must vanish for any such observable $A$ and any
equilibrium ensemble $\rho_{\rm eq}$ in Eq.~(\ref{15}).

The same line of reasoning can be readily extended to 
arbitrary observables of the general structure
$A=\sum_\mu A_\mu$,  where each $A_\mu$ consists of a product of factors 
of the form $s_l^a$ so that the number of factors with $a\in\{y,z\}$ is {\em odd}.

Since $H$ commutes with $U$ (see Eq.~(\ref{b23})), 
there exists a common set of eigenvectors $|n\rangle$,
and since $U$ is unitary, all its eigenvalues are of unit modulus,
implying $\langle n|U^\dagger A U|n\rangle=\langle n|A|n\rangle$
for any Hermitian operator $A$.
Choosing $A=s_l^z$,
we moreover can conclude from (\ref{b21})
that $\langle n|U^\dagger A U|n\rangle=-\langle n|A|n\rangle$ and
thus $\langle n| A |n\rangle=0$ for all $n$ and all $l$,
and likewise for the more general $A$'s from
the previous paragraph.

In other words, there exists at least one energy eigenbasis 
for which even the strong ETH is trivially fulfilled.
Moreover, this property is simultaneously true
for all the above specified observables $A$.
We also recall that all so far conclusions do not require 
translational invariance of $H$ (see below Eq.~(\ref{b17})).
On the other hand, if $H$ does commute with the translation operator
$T$, and since also $U$ commutes with $T$, we can even choose 
a common eigenbasis of $H$, $U$, and $T$.
In other words, there must exist even a translationally invariant 
basis for which  the strong ETH is trivially fulfilled 
for a large class of observables.
On the other hand, it also must be emphasized that the conventional
understanding of the (strong and weak) ETH is that
its claims must be fulfilled by {\em all} local observables.

As an example we finally turn to the spin-1 Hamiltonian from Eq.~(\ref{55}).
The remaining task is to show that this Hamiltonian is as 
specified in and below Eq. (\ref{b17}).
Without loss of generality, we focus on the first summand with $l=1$
on the right-hand side of (\ref{55}),
\begin{eqnarray}
Q:=s_1^+(s_{2}^-)^2s_{3}^+ 
\ .
\label{b27}
\end{eqnarray}
Recalling that $s_l^\pm:=s_l^x\pm is_l^y$
implies that the last two factors commute and thus
\begin{eqnarray}
Q & = & [(s_1^x+is_1^y)(s_3^x+is_3^y)] [s_2^x-is_2^y]^2 
\nonumber
\\
& = & [B_1+iB_2][B_3-iB_4]
\ ,
\label{b28}
\\
B_1 & := & s_1^x s_3^x - s_1^y s_3^y
\ ,
\label{b29}
\\
B_2 & := & s_1^x s_3^y + s_1^y s_3^x
\ ,
\label{b30}
\\
B_3 & := & (s_2^x)^2-(s_2^y)^2
\ ,
\label{b31}
\\
B_4 & := & s_2^xs_2^y+s_2^ys_2^x
\ .
\label{b32}
\end{eqnarray}
It
follows
that
\begin{eqnarray}
Q & = & Q_1 + i Q_2
\ ,
\label{b33}
\\
Q_1 & := & B_1 B_3 + B_2 B_4 
\ ,
\label{b34}
\\
Q_2 & := & B_2 B_3  - B_1 B_4 
\ .
\label{b35}
\end{eqnarray}
One readily verifies that the operators $B_1,...,B_4$
are Hermitian and that all products appearing on 
the right-hand side of (\ref{b34}) and (\ref{b35}) consist of 
commuting factors.
It follows that also $Q_1$ and $Q_2$ are Hermitian 
and therefore $Q+Q^\dagger=2 Q_1$.
Moreover, for both summands contributing to $Q_1$ in (\ref{b34}),
the number of factors $s_l^a$ with the property 
$a\in\{y,z\}$  is even.
Hence the same must indeed be true for each summand 
appearing on the right-hand side of Eq.~(\ref{55}).

We finally remark that somewhat similar symmetry properties 
are also mentioned (but not derived) in Appendix A 
of Ref.~\cite{sal20},  see in and around Eq. (A2) therein.

\section{Numerical evaluation of the ETH quantifier via dynamical typicality}
\label{app:DynTyp}

In this appendix,
we explain how we evaluated the ETH quantifier $\Delta_{A,\eq}^2(\beta)$ from~\eqref{54} 
numerically for the two basis choices in Fig.~\ref{fig:ETHVio}:
the maximally ETH-violating basis and a translationally invariant basis.

We adopt standard dynamical-typicality methods \cite{bar09, rei20} to calculate canonical expectation values and correlation functions in high-dimensional Hilbert spaces,
exploiting the symmetries of the problem.
Concretely,
the Hamiltonian~\eqref{55} commutes with the translation operator $T$ from~\eqref{4},
the total magnetization
\begin{equation}
\label{eq:Mz}
	M^z := \sum_{l=1}^L s_l^z \,,
\end{equation}
and the dipole moment
\begin{equation}
	P^z := \sum_{l=1}^L (l - l_0) s_l^z \,,
\end{equation}
where $l_0$ is a fixed reference site.
The magnetization commutes with both $P^z$ and $T$,
but $[P^z, T] \neq 0$,
so we can use the eigenvalues of either $(M^z, P^z)$ or $(M^z, T)$ to label symmetry sectors by good quantum numbers.
For the numerics in Sec.~\ref{s5},
we choose $(M^z, P^z)$ because the product basis~\eqref{3} is an eigenbasis for this pair,
making it particularly convenient to define the associated subspaces.
The magnetization $M^z$ has $2 L + 1$ distinct eigenvalues $m \in \{-L, -L+1, \ldots, L \}$.
Since the values of the dipole moment must be invariant under shifts of the reference site $l_0$ by multiples of $L$,
the eigenvalues of $P^z$ are defined modulo $L$,
e.g., $p \in \{ 0, \ldots, L - 1 \}$.

If $q = (m, p)$ denotes the (pair of) quantum numbers identifying a symmetry sector,
we define $P_q$ as the projector onto this sector and
\begin{equation}
	\tr_q A := \tr(P_q A)
\end{equation}
as the trace of an arbitrary observable $A$ within the sector.
In particular, $N_q := \tr_q \id$ is the sector dimension (number of states in the sector).
Note that some of the so-defined sectors [e.g., $(\pm L, p)$ with $p > 0$]
are empty ($N_q = 0$).

We can now decompose canonical expectation values into contributions from the symmetry sectors,
\begin{equation}
\label{eq:Acan:trq}
	\< A \>_{\!\rho_{\beta}} := Z_\beta^{-1} \tr(e^{-\beta H} A)
		= Z_\beta^{-1} \sum_q \tr_q(e^{-\beta H} A) \,,
\end{equation}
where
\begin{align}
\label{eq:Zbeta}
	Z_\beta &:= \tr e^{-\beta H} = \sum\nolimits_q Z_{\beta, q} \,, \\
\label{eq:Zbeta,q}
	Z_{\beta, q} &:= \tr_q e^{-\beta H} \,.
\end{align}
Within every symmetry sector,
we employ dynamical typicality as usual to approximate traces by expectation values of Haar-random pure states:
If $\ket{\phi_q}$ is a normalized Haar-random state in the sector $q$,
then
\begin{equation}
	\mathbb{P}(\lvert \bra{\phi_q} A \ket{\phi_q} - \tr_q A / N_q \rvert \geq \varepsilon) \leq \frac{\lVert A \rVert_{\mathrm{op}}^2}{\varepsilon^2 N_q} \,,
\end{equation}
hence $\tr_q A \simeq N_q \bra{\phi_q} A \ket{\phi_q}$ if $N_q \gg 1$ \cite{llo88, gem04}.

Defining
\begin{equation}
	\ket{\phi_{\beta,q}} := e^{-\beta H / 2} \ket{\phi_q}
\end{equation}
and recalling Eqs.~\eqref{eq:Acan:trq}-\eqref{eq:Zbeta,q},
we can calculate
\begin{equation}
\label{eq:Acan:typq}
	\< A \>_{\!\rho_\beta}
		\simeq \frac{ \sum_q N_q \, \bra{\phi_{\beta,q}} A \ket{\phi_{\beta,q}} }{ \sum_q N_q \, \< \phi_{\beta,q} | \phi_{\beta,q} \> } \,.
\end{equation}
To calculate correlation functions $\< A \, B(t) \>_{\!\rho_{\beta}}$,
where $B(t) := e^{i H t} B e^{-i H t}$ is the time-evolved observable $B$ in the Heisenberg picture,
we furthermore introduce
\begin{equation}
\label{eq:phiA}
	\ket{\phi_{\beta,q}^A} := A \ket{\phi_{\beta,q}}
\end{equation}
and
\begin{equation}
\label{eq:phi(t)}
	\ket{\phi_{\beta,q}(t)} := e^{-i H t} \ket{\phi_{\beta,q}} \,, \quad
	\ket{\phi_{\beta,q}^A(t)} := e^{-i H t} \ket{\phi_{\beta,q}^A} \,.
\end{equation}
Replacing $A$ by $A\, B(t)$ in~\eqref{eq:Acan:typq}
and adopting Eqs.~\eqref{eq:phiA} and~\eqref{eq:phi(t)},
we obtain
\begin{equation}
\label{eq:ABcorr:typq}
	\< A \, B(t) \>_{\!\rho_\beta}
		\simeq \frac{ \sum_q N_q \, \bra{\phi_{\beta,q}^A(t)} B \ket{\phi_{\beta,q}(t)} }{ \sum_q N_q \, \< \phi_{\beta,q} | \phi_{\beta,q} \> } \,.
\end{equation}

This leaves us with a recipe
to calculate canonical expectation values:
We
\begin{enumerate}
	\item generate a Haar-random pure state $\ket{\phi_q}$
		in every symmetry sector $q$;
	\item calculate all $\ket{\phi_{\beta,q}}$ by imaginary-time Suzuki-Trotter propagation of $\ket{\phi_q}$ with Hamiltonian $H$ from imaginary time $0$ to $\frac{\beta}{2}$;
	\item calculate all $\bra{\phi_{\beta,q}} A \ket{\phi_{\beta,q}}$ and $\< \phi_{\beta,q} | \phi_{\beta,q} \>$;
	\item combine the results from all sectors according to~\eqref{eq:Acan:typq} to obtain $\< A \>_{\!\rho_\beta}$.
\end{enumerate}
To calculate correlation functions,
we proceed and
\begin{enumerate}[resume]
	\item calculate $\ket{\phi_{\beta,q}^A}$ according to~\eqref{eq:phiA} for every sector $q$ and inverse temperature $\beta$ of interest;
	\item calculate all $\ket{\phi_{\beta,q}(t)}$ and $\ket{\phi_{\beta,q}^A(t)}$ by real-time Suzuki-Trotter propagation of $\ket{\phi_{\beta,q}}$ and $\ket{\phi_{\beta,q}^A}$, respectively, with Hamiltonian $H$ from time $0$ to $t$;
	\item calculate all $\bra{\phi_{\beta,q}^A(t)} B \ket{\phi_{\beta,q}(t)}$;
	\item combine the results from all sectors according to~\eqref{eq:ABcorr:typq} to obtain $\< A \, B(t) \>_{\!\rho_\beta}$.
\end{enumerate}

To increase and estimate the accuracy,
we repeat this procedure for eight (four) independent sets of Haar-random states $\ket{\phi_q}$
for the systems with $L \leq 11$ ($L > 11$) in Figs.~\ref{fig:ETHVio} and~\ref{fig:perturbations}
and calculate the standard error (shown as error bars).
In Fig.~\ref{fig:ETHVioSubsectors}, the Hilbert space dimensions are smaller because we inspect symmetry sectors, hence we used larger sets of $1000$, $200$, and $16$ Haar-random states, respectively, for the systems with $L = 9$, $11$, and $13$.

Turning to the ETH quantifier $\Delta_{A,\eq}^2(\beta)$,
we first inspect the maximizing energy eigenbasis,
i.e.,
the basis in which $A$ is diagonal in every degenerate subspace.
As derived in Sec.~\ref{s62},
$\Delta_{A,\eq}^2(\beta)$ is equal to the time-averaged auto-correlation function
$\overline{C_{\! A}(t)}$ in this case,
cf.\ Eq.~\eqref{25}, in particular.
Explicitly, we thus have
\begin{equation}
\label{eq:ETHVio:MaxBasis}
	\Delta_{A,\eq}^2(\beta)
		= \overline{ \< A \, A(t) \>_{\!\rho_\beta} } - \< A \>_{\!\rho_\beta}^2 \,.
\end{equation}
Following the recipe outline above,
we compute $\< A\,A(t) \>_{\!\rho_\beta}$ and $\< A \>_{\!\rho_\beta}$ via dynamical typicality,
and time-average the former in the interval $t \in [10, 20]$
to eventually obtain $\Delta_{A,\eq}^2(\beta)$ from~\eqref{eq:ETHVio:MaxBasis}.

To evaluate $\Delta_{A,\eq}^2(\beta)$ for a translationally invariant basis $\{ \ket{n} \}$,
we observe that the diagonal matrix elements $A_{nn}$ of the local observable $A$
coincide with those of the corresponding intensive observable $B := \frac{1}{L} \sum_{l = 1}^L T^l A T^{-l}$, i.e., $A_{nn} = B_{nn}$.
Focusing on the special case $A = s_1^z$ of interest in Fig.~\ref{fig:ETHVio},
we furthermore recall that $A_{\mathrm{eq}} = 0$.
We also observe that $B = M^z$ is the total $z$ magnetization from~\eqref{eq:Mz},
which commutes with $H$ from~\eqref{55} and $T$ from~\eqref{4},
meaning that we can choose the specific $T$-invariant energy eigenbasis $\{ \ket{n} \}$
such that $B$ is diagonal, too.
The ETH quantifier $\Delta_{A,\eq}^2(\beta)$ can then be expressed in this basis as
\begin{equation}
	\Delta_{A,\eq}^2(\beta) = \sum_n p_n A_{nn}^2 = \sum_n p_n B_{nn}^2 = \< B^2 \>_{\!\rho_{\beta}} \,,
\end{equation}
where we used $A_{\mathrm{eq}} = 0$ in the first equality
and the fact that $B$ is diagonal in the chosen $T$-invariant basis in the last equality.
Since the last representation is just a canonical expectation value,
we can evaluate it as before using dynamical typicality and imaginary time evolution
in the $(M^z, P^z)$ symmetry sectors.



\begin{thebibliography}{770}
\bibitem{note:EquilibriumState}
Since a local quench cannot noticeably change the expectation values of 
extensive conserved quantities in the thermodynamic limit,
the pertinent equilibrium state that would describe the long-term properties 
of a thermalizing system is still $\rho_{\eq}$ from Eq.~\eqref{42}
with the same inverse temperature $\beta$ as in Eq.~\eqref{57} (see also Refs.~\cite{far17, dab22}).
\end{thebibliography}
\end{document}